# A Molecular Dynamics Study on $CO_2$ Diffusion Coefficient in Saline Water Under a Wide Range of Temperatures, Pressures, and Salinity Concentrations: Implications to $CO_2$ Geological Storage

Sina Omrani[a†], Mehdi Ghasemi[b†], Saeed Mahmoodpour[c*], Ali Shafiei[b*], Behzad Rostami[a]

[a] *Institute of Petroleum Engineering, College of Engineering, University of Tehran, Tehran, Iran.*

[b] *Petroleum Engineering Program, School of Mining & Geosciences, Nazarbayev University, Nur-Sultan, Astana 010000, Kazakhstan*

[c] *Institute of Applied Geosciences, Geothermal Science and Technology, Technische Universität Darmstadt, Darmstadt, Germany*

[†] *S. Omrani and M.Ghasemi contributed equally to this work.*

** Corresponding authors*

*E-mail addresses: saeed.mahmoodpour@tu-darmstadt.de (S. Mahmoodpour), ali.shafiei@nu.kz.edu (A. Shafiei)*

# Highlights

- The $CO_2$ diffusion coefficient monotonically decreases with increasing NaCl concentration.
- At higher temperatures, the $CO_2$ diffusion coefficient degree of reduction is more severe.
- Pressure has a relatively negligible impact on the $CO_2$ diffusion coefficient in water/brine.

# Graphical Abstract

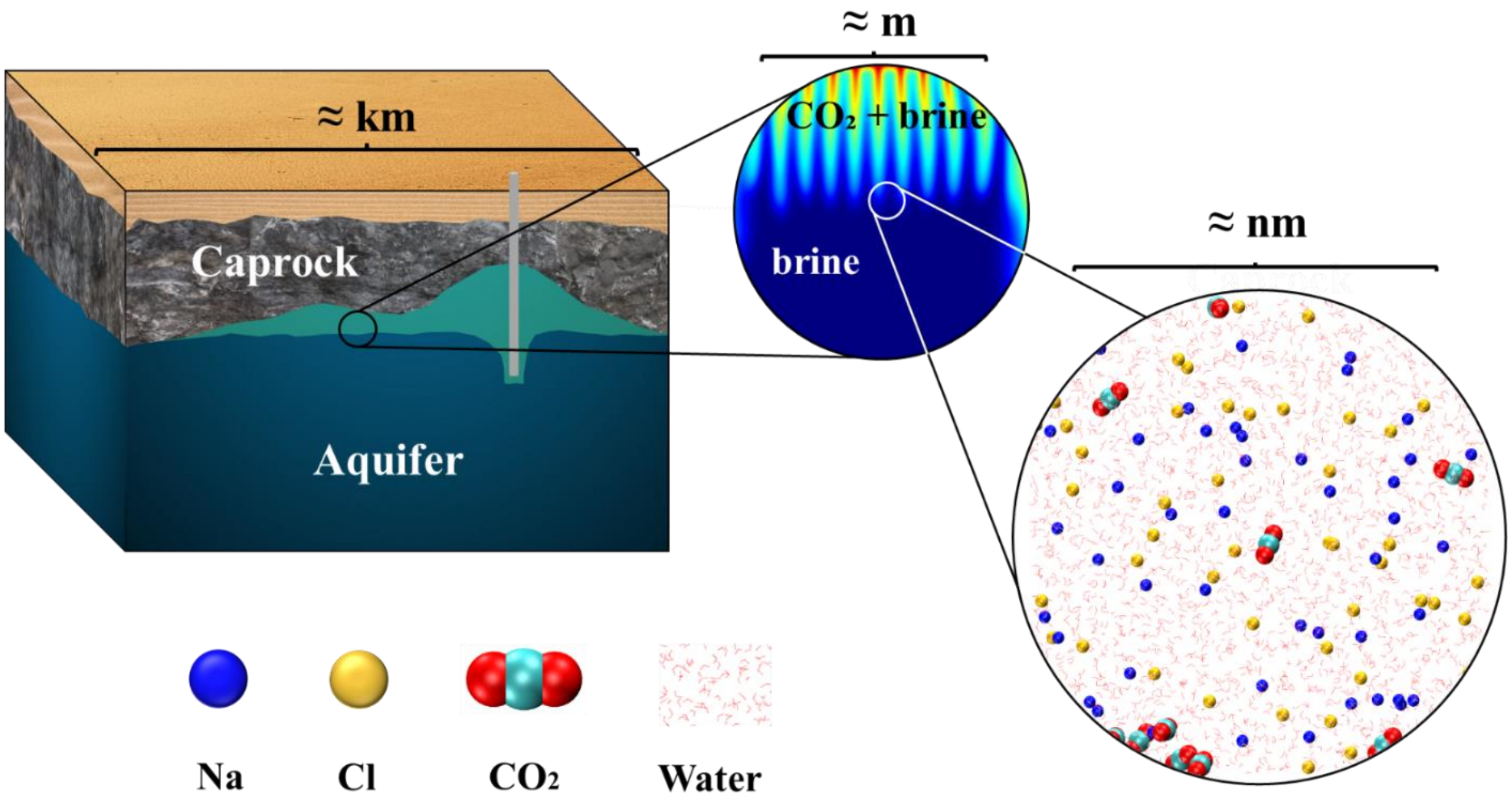

# Abstract

Carbon dioxide ($CO_2$) sequestration in saline aquifers has been introduced as one of the most practical, long-term, and safe solutions to tackle a growing threat originating from the emission of $CO_2$. Successfully executing and planning the process necessitates a comprehensive understanding of $CO_2$ transport properties- particularly the diffusion coefficient, influencing the behavior of $CO_2$ dissolution in water/brine regarding the shape of viscous fingers, the onset of instabilities, etc. In this research, Molecular Dynamics (MD) simulation was employed to compute the $CO_2$ diffusion coefficient in various NaCl saline water concentrations (1-6 mol NaCl/kg water) under the broad spectrum of temperatures (294-423 K) and pressures (10-30 MPa) to acquire a data-set for the $CO_2$ diffusion coefficient in different circumstances. According to the results, the NaCl concentration increase gives rise to a decrease in the $CO_2$ diffusion coefficient, by which the reduction is most notably at higher temperatures. In fact, the presence of more cations due to the salinity concentration increase forms further hydration shells, acting as a barrier for $CO_2$ diffusion. In addition, the rise in the $CO_2$ diffusion at elevated temperatures can be explained by the cation's hydration shell size reduction with temperature increment due to intensifying repulsive forces among water molecules. We also proposed a new precise correlation for estimating $CO_2$ diffusion coefficients over temperatures and salinity ranges of this study. Regarding the pressure variation effects, no tangible changes are observed with pressure increase, validating a negligible influence on the diffusion coefficient. Furthermore, the variability of the $CO_2$ diffusion coefficient in the presence of other salts, namely $MgCl_2$, $CaCl_2$, KCl, and $Na_2SO_4$, were computed separately. Comparing the influence of various salts, $CaCl_2$ and KCl have the highest and lowest effect on the $CO_2$ diffusion coefficient, respectively. Finally, a set of direct numerical simulations was conducted to study the impact of the $CO_2$ diffusion coefficient on the $CO_2$ dissolution process. The results shed light on the importance of $CO_2$ diffusion coefficient changes under the saline water condition in predicting dissolution process behavior and further calculations.

# 1. Introduction

The increase of greenhouse gases, in particular carbon dioxide ($CO_2$), has become a severe challenge that puts the life of most species in danger. To minimize $CO_2$ emission in the environment, a promising solution of $CO_2$ storage in saline aquifers has been introduced as a viable option due to the high capacity, chemistry, and geological accessibility of saline aquifers [1-3]. In order to have long-term safety of underground $CO_2$ storage, a thorough assessment of trapping mechanisms in the saline aquifers is essential [4]. In this regard, the $CO_2$ dissolution mechanism is considered a necessity among trapping mechanisms to bring about enduring $CO_2$ storage [5]. The dissolution process initiates with $CO_2$ diffusion into brine. The relatively higher density of the newly formed diffuse layer of $CO_2$ and brine on top of an aquifer compared to the original brine (i.e., brine without the dissolved $CO_2$) results in gravitational instabilities. These instabilities grow with time and eventuate to the convective fingers, which gradually move downward. The replacement with fresh brine is continued until the aquifer reaches its maximum dissolution capacity and can no longer dissolve more $CO_2$. Dissolution of $CO_2$ from one aspect prepares the conditions for the permanent storage of the $CO_2$ through mineralization. From the other aspect, it decreases the stress on the caprock through pressure reduction. Furthermore, dissolved $CO_2$ has a higher viscosity and density compared to the gas phase, and the possibility of leakage would decrease. Therefore, the higher $CO_2$ dissolution rate into the brine reduces leakage risk.

The behavior of the dissolution process in terms of instability onset, the shape and number of convective fingers, and dissolution flux is considerably affected by the diffusion coefficient and its controlling parameters such as temperature, pressure, salinity, etc., during the whole diffusion-convection process. The role of salinity or impurities in $CO_2$ diffusion coefficient changes has been ignored in previous works, in which either the $CO_2$ diffusion coefficient in pure water was used [6-9], the works have been limited to a range of available data [10], or they tried to calculate effective diffusion coefficient based on experimental results [11]. Nevertheless, limited studies have attempted to assess the $CO_2$ diffusion coefficient in water in the presence of various impurities and salts [12-18]. A summary of the literature is provided hereunder.

In earlier research regarding $CO_2$ storage in an aquifer, measurement of the $CO_2$ diffusion in pure water under a short range of temperature and pressure was the center of attention [19-25]. Nevertheless, over time, studies have used better techniques to measure $CO_2$ diffusion coefficients. In 1994, Tamimi et al. [26] used a wetted-sphere absorption apparatus to measure the diffusion coefficients for $H_2S$, $CO_2$, and $N_2O$ in water at the temperature range of 293-368 K. Hirai et al. [27] predicted the $CO_2$ diffusion coefficient by measuring the liquid $CO_2$ droplet-dissolution rate. In 2012, Sell et al. [15] used a microfluidic-based approach to measure the $CO_2$ diffusion coefficient in water and NaCl brine. They measured $CO_2$ diffusion in brine at 299 K, 0.5 MPa, and up to 5 molars (M) NaCl concentration and proposed an exponential relationship between the diffusion coefficient and salinity. In 2014, Cadogan et al. [28] measured $CO_2$ and $N_2$ diffusion coefficients for a broad spectrum of temperatures and pressure via the Taylor dispersion method. In 2015, their research team measured the $CO_2$ diffusion coefficient in brine at 298 K and 0.1 bar MPa using Nuclear Magnetic Resonance (NMR) method. Several types of salts were also investigated (NaCl, $CaCl_2$, and $Na_2SO_4$) [29]. Zhang et al. [16] used the pressure decay method to measure the $CO_2$ diffusion coefficient. They evaluated the effect of temperature, pressure, salt concentration, and type of salt on the $CO_2$ diffusion coefficient. Their study showed that the diffusion coefficient increases by increasing temperature or pressure while decreasing with salinity. Belgodere et al. [17] applied Raman spectroscopy to compute the $CO_2$ diffusion coefficient in pressurized aqueous solution (both pure and saline water) at room temperature. They also claimed the inverse relationship of the $CO_2$ diffusion coefficient with salinity. Jafari Raad et al. [4] also measured the $CO_2$ diffusion coefficient in pure and saline water under an extensive range of temperature and pressures by employing the pressure decay method. The Supporting Information (SI) provides a list of previous works done on this subject in **Table S1**.

The experimental measurement of the diffusion coefficient entails a complex and time-consuming procedure. Therefore, finding a great alternative technique is vital. With the advent of computational methodologies, Molecular Dynamics (MD) simulation has been demonstrated as a valuable method for evaluating physicochemical characteristics and fluids' thermodynamic and transport properties [30-32]. Heretofore, it has been established that MD simulations can be used to calculate diffusion coefficients [33, 34]. For example, Garcia-Rates et al. [35] employed MD simulation to estimate the $CO_2$ diffusion coefficient in NaCl brine under the influence of temperature (333.15, 393.15, and 453.15 K), pressure (10 and 20 MPa), and salinity (1,2, and 4 M). They observed that an increase in the temperature causes the diffusion coefficient increment, while no accurate relationship is present for the dependency of the diffusion coefficient on salinity or $CO_2$ concentration. Moultos et al. [36] estimated the $CO_2$ diffusion coefficient via MD simulations, as well. They compared diverse combinations of force fields for $CO_2$ (EPM2 and TraPPE) and water (SPC, SPC/E, and TIP4P/2005) by considering various temperatures and pressures. Finally, they proposed a correlation for predicting the $CO_2$ diffusion coefficient in water.

To the best of the authors' knowledge, although some researchers have emphasized the impact of salinity on the diffusion coefficient variation, the literature lacks a comprehensive set of $CO_2$ diffusion coefficient data in saline water in a comprehensive range of temperatures, pressures, and salt concentrations. In this research, we aimed to apply MD simulations to estimate the $CO_2$ diffusion coefficient at the temperature and pressure range of 294-423 K and 10-30 MPa, respectively, under NaCl concentrations of 1, 2, 4, and 6 M. Further investigation was conducted to scrutinize the effect of 4 different types of salt, including $MgCl_2$, $CaCl_2$, KCl, and $Na_2SO_4$, on the diffusion coefficient, and then compared with the NaCl. Next, according to the obtained values, we proposed a novel correlation to predict the $CO_2$ diffusion coefficient in saline brine, which can be valuable in engineering applications. In the final stage, to clarify the influence of ignoring the dependency of the $CO_2$ diffusion coefficient on salinity and assuming the $CO_2$ diffusion coefficient value in pure water instead, we analyzed the $CO_2$ dissolution behavior using direct numerical simulation at a temperature of 323 K. The primary aim of the study is to provide some reliable data, which can serve as a reference, to enhance the accuracy of Carbon Capture and Sequestration/Storage (CCS) planning and executing.

The remainder of this paper is organized as follows. First, we described the theory of diffusion coefficient calculation and details of the MD simulations in Section 2. Then, results and in-depth interpretation of analyses were provided in Section 3. Finally, the conclusions of the research were drawn in section 4.

# 2. Methodology

In this section, we first described the different diffusion coefficient types and their calculation methods via MD simulation results, and then we discussed details of the simulation procedure.

## 2.1. Diffusion and Related Theories

Mass transfer computation requires an accurate description of the diffusion coefficient and thus diffusion mechanism. Diffusion coefficients are classified as follows: 1) Self-diffusion coefficient ($D_{Self}$), which describes the transport of particles due to Brownian motion, 2) Maxwell-Stefan (MS) diffusion coefficient ($D_{MS}$) that represents friction between two components, and 3) Fick diffusion coefficient ($D_{Fick}$), which is resulting from the concentration gradient of a component [12]. The diffusion coefficient of different kinds can be studied and validated with two different MD simulation methods: equilibrium MD (EMD) and non-equilibrium MD (NEMD). There is a difference between these two methods in whether the external force is present or not. Generally, due to some problems existing in the NEMD method, the EMD method, which is based on analyzing the equilibrium fluctuations, has been more widely adopted to calculate the diffusion coefficient [31, 37, 38]. Calculating the $D_{Self}$, as a concern of this study, using MD can be predicted following two methods: Einstein and Green-Kubo [39]. According to the literature, the Einstein method is more widely preferred [12], and it is defined as follows:

$$D_{i,self} = \frac{1}{6N_i} \lim_{m \to \infty} \frac{1}{m.\Delta t} \langle \sum_{l=1}^{N_i} \left( r_{l,i}(t + m.\Delta t) - r_{l,i}(t) \right)^2 \rangle \tag{1}$$

where $N_i$ is the number of particles in component i, m is the number of time steps, $\Delta t$ is time step used in simulation, and $r_{l,i}(t)$ is the position of molecules $l^{th}$ of component at time t.

Since the concept of collective diffusivities ($D_{MS}$ and $D_{Fick}$) is more interesting in engineering applications, a component's net transport in a multicomponent mixture can be explained by means of collective diffusion [34]. In spite of $D_{Fick}$, $D_{MS}$ cannot be measured experimentally, and the thermodynamic correction factor ($\Gamma$) relates these two theories (see Eq. (*2*) or a binary system). In an infinitely diluted system (which $\Gamma$ is almost unity), all of these values are approximately equal [28, 40].

$$[D_{Fick}] = [D_{MS}][\Gamma] \quad (2)$$

According to the study conducted by Garcia-Rates et al. [35], a $CO_2$ brine system is thermodynamically ideal. As well, we calculated the Γ for some of our systems and have found similar results. We refer you to our previous paper for more details on the diffusion coefficient types and how to calculate them using MD [12].

## 2.2. Simulation Details

### 2.2.1. Molecular Dynamics (MD) Simulation

MD simulation technique, as a powerful gadget, can be employed to have a better understanding of the behavior and characteristics of simple and/or complex systems from the atomic level, particularly calculating the $CO_2$ diffusion coefficient in this study. Since the accuracy of MD simulation results is strongly dependent on the applied intra-intermolecular potentials, first, we conducted comparative MD simulations to determine the most precise force field for the $CO_2$ at infinite dilution in the saline brine system. Thus, three different developed force fields for $CO_2$, namely rigid and flexible TraPPE models and the EPM2 model [41-43], and two salt models of Smith and Dang and Joung and Cheatham (we refer to them as Smith and JC models, respectively) [44, 45] in combination with the SPCE model [46] for water molecules were employed to identify the combined potentials with the highest accuracy. The basis of comparison was the available $CO_2$ diffusion coefficient data in the literature obtained by Belgodere et al. [17] via the experimental measurement. According to the several statistical analyses [47], including average percent relative error (APRE), average percent absolute relative error (AAPRE), standard deviation (SD) of error, and root mean square error (RSME) and coefficient of determination ($R^2$), the combined TraPPE (Flex)-JC-SPCE potentials showed the best fit with the literature data for almost all cases, and it was selected for the rest of the simulations. Note that the SI provides the results of statistical analyses in more detail.

All MD simulations were performed by the GROMACS package (version 2021.1) [48]. A cubic box with the size of 4.5×4.5×4.5 $nm^3$ was selected, and the periodic boundary condition (PBC) was applied in all directions. Each system followed three simulation steps: initially, the energy minimization using steepest descent algorithm to reduce bad contacts and produce stable configuration for simulation, then the constant isothermal-isobaric ensemble (NPT) run for 3 ns to reach the desired temperature and pressure, and finally, the canonical ensemble (NVT) simulation for 60 ns to collect the data. A Nosé–Hoover thermostats and a Parrinello–Rahman barostat were applied to maintain the temperature and pressure, respectively [49-51]. The long-range electrostatic interactions were calculated using the particle-mesh Ewald (PME) summation technique with a precision of $10^{-4}$. A cut-off radius of 1.2 nm was assigned for both van der Waals (vdW) and electrostatic interactions. Also, all hydrogen bonds were constrained by the Linear Constraint Solver (LINCS) algorithm [52]. Regarding the intermolecular interactions, for all cases except EPM2, which originally used geometric mean, the interactions between molecules were computed using Lorenz-Berthelot mixing rules [53]. It is worth mentioning that all simulations were repeated three times to enhance the accuracy of the results.

Note that since the finite-size simulation box impacts the diffusion coefficient calculated by MD simulations [34, 54], two other sizes of 3 and 7 nm were considered to track the changes in diffusion coefficient values (see **Figure S1**). According to the results, although the variation in the diffusion coefficients was relatively negligible, we corrected even the small effect (almost 2 orders of magnitude lower than the diffusion coefficient) by the Yeh–Hummer method [55].

$$D_{Self} = D_{Self}^{MD} + \frac{k_B T \xi}{6\pi\eta L} \tag{3}$$

where $k_B$ and T are the Boltzmann constant and temperature, respectively. $\xi$ is a dimensionless constant equal to 2.837298 for periodic cubic boxes, $\eta$ is the shear viscosity, and L is the box size in nm.

## 2.2.2. Direct Numerical Simulation

To clarify the dependency of $CO_2$ dissolution behavior on the diffusion coefficient in saline brine, we applied direct numerical simulation based on the finite element method using COMSOL software. We chose cases at 323 K and 10 MPa. To capture the dissolution behavior, the following equations as the mass transfer (Eq. (*4*)) and continuity (Eq. 5a) equations should be solved in a fully coupled scheme:

$$\phi \frac{\partial C}{\partial t} = \phi \nabla . (D \nabla C) - V . \nabla C \quad (4)$$

$$\nabla U = 0 \quad (5a)$$

$$U = -\frac{k}{\mu}(\nabla P - \rho g) \quad (5b)$$

where ϕ, C, D, U, k, P, μ, ρ, and g represent porosity, dissolved $CO_2$ concentration, diffusion coefficient, the Darcy velocity vector, permeability, pressure, the viscosity, the density, and the gravity acceleration constant, respectively. Duan and Sun paper was used to get the solubility of $CO_2$ in water/brine [56]. We applied the Garcia method to calculate the $CO_2$-rich brine density [57], and the density and viscosity of brine were computed by Mao and Daun [58, 59]. We used an equation from our previous work to estimate the onset of the shut-down regime to determine how long each simulation should be continued [60]. Our previous papers entail more detail on the direct numerical simulation [11, 61]. These equations are solved once with the diffusion coefficients calculated from this study and again with considering the $CO_2$ diffusion coefficient in pure water in order to demonstrate their impact on $CO_2$ dissolution process behavior.

# 3. Results and Discussion

In this section, the $CO_2$ diffusion coefficient values in brine under various conditions are calculated. The impact of four factors, namely NaCl salinity concentration, temperature, pressure, and salinity type, were scrutinized. The radial distribution function (RDF) and H-bonds analyses were also discussed to elucidate the reason behind the results.

## 3.1. $CO_2$ Diffusion Coefficient Under Various NaCl Salinities and Temperatures

The combined effects of temperature and salinity alteration on the $CO_2$ diffusion coefficient are shown in **Figure *1***. In order to better visualize the trends, we categorized the results into relatively low and high temperatures (**Figure *1*a** and **b**, respectively). Generally, there is a similar consensus over the impacts of temperature and salinity on the diffusion coefficient. A rise in temperature results in greater kinematic energy and, therefore, a higher diffusion coefficient, while an increase in NaCl salinity results in a lower diffusion coefficient. In more detail, the salinity increment brings about a monotonical reduction of the diffusion coefficient, by which the reduction is more severe at the beginning of curves for higher temperatures. Compared to pure water, the $CO_2$ diffusion coefficient in brine is reduced, on average, by 15%, 29%, 49%, and 64%, respectively, at 1 M, 2 M, 4 M, and 6 M solutions. Since the trend follows an exponential behavior at each temperature, a regression analysis was performed using an optimization code in Python to provide a correlation between the $CO_2$ diffusion coefficients in an aqueous solution at temperatures ranging from 294 to 423 K and NaCl concentrations of 0 to 6 M.

$$\text{Diffusion Coefficient} = -18.157948 \times e^{-0.05736C} + 0.068700205361 \times T - 0.068700205361 \times C^{0.820561458} \times T^{1.46331515077} \quad (6)$$

where C is the molarity of NaCl and T is the temperature in K. All results of this study are presented in **Table *1***. Also, to get the temperature dependence of the $CO_2$ diffusion coefficient, the Arrhenius law was used.

$$D = D_0 e^{-\frac{E_a}{RT}} \quad \text{or} \quad \ln D = \ln D_0 - \frac{E_a}{RT} \quad (7)$$

where D, $D_0$, $E_a$, R, and T are the diffusion coefficient, the pre-exponential factor, the activation energy, the gas constant, and the absolute temperature. **Figure *2*** depicts the Arrhenius fits of the results of this study. By calculating their slope and putting it in equation (*7*), we can evaluate the activation energy for each of them. In the **Figure 2**, it is evident that the $E_a$value is almost constant up to 2 M. However, as the salinity increases, the value of $E_a$ also increases, which means the diffusion mechanisms become harder and slower.

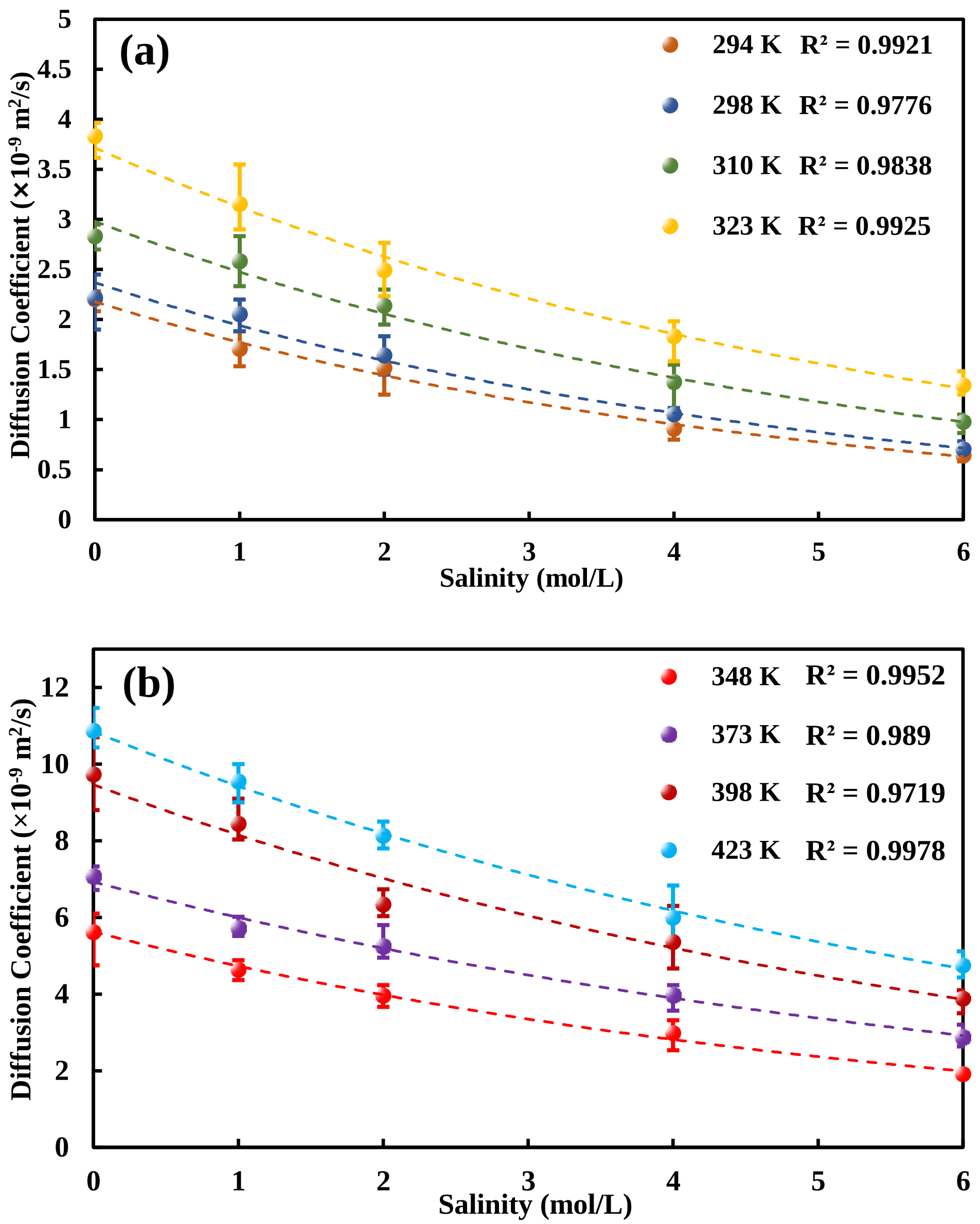

**Figure 1**- The $CO_2$ diffusion coefficient in 0-6 M NaCl solution at 10 MPa and temperatures of: a) 294-323 K b) 348-423 K.

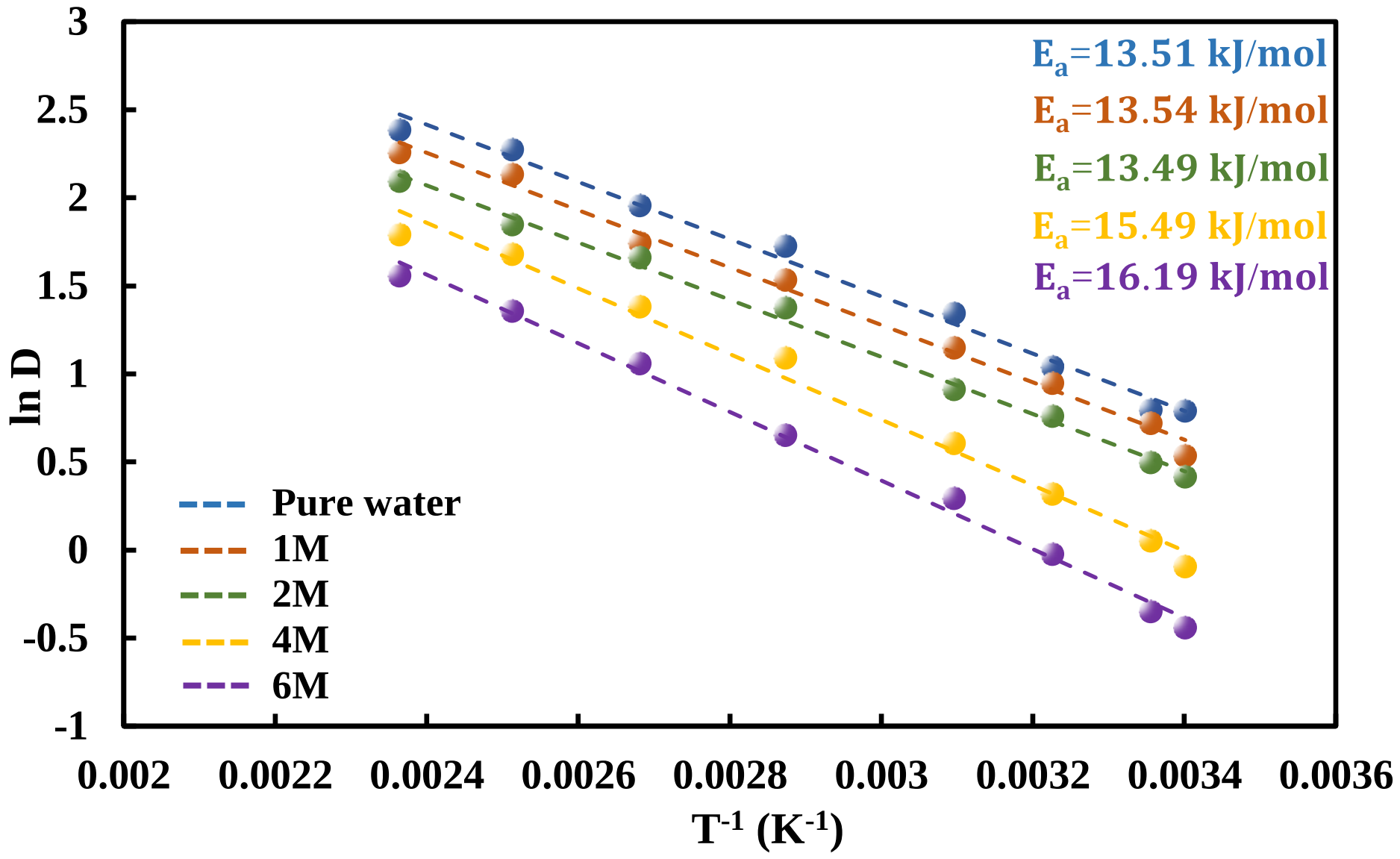

**Figure 2**- The Arrhenius plot of the $CO_2$ diffusion coefficient in water/brine for different NaCl concentrations.

**Table 1**- The results of the $CO_2$ diffusion coefficient in pure water/NaCl brine under various temperatures and pressures.

| No. | Case Name | Solution $\left(\frac{mol}{L}\right)$ | Temperature (K) | Pressure (MPa) | D $(\times 10^{-9}\frac{m^2}{s})$ |
|---|---|---|---|---|---|
| 1 | $T_{294}C_0$ | Pure Water | 294 | 100 | 2.2053 |
| 2 | $T_{294}C_1$ | 1 M NaCl | 294 | 100 | 1.7082 |
| 3 | $T_{294}C_2$ | 2 M NaCl | 294 | 100 | 1.5163 |
| 4 | $T_{294}C_4$ | 4 M NaCl | 294 | 100 | 0.9109 |
| 5 | $T_{294}C_6$ | 6 M NaCl | 294 | 100 | 0.6443 |
| 6 | $T_{298}C_0$ | Pure Water | 298 | 100 | 2.2167 |
| 7 | $T_{298}C_1$ | 1 M NaCl | 298 | 100 | 2.0553 |
| 8 | $T_{298}C_2$ | 2 M NaCl | 298 | 100 | 1.6443 |
| 9 | $T_{298}C_4$ | 4 M NaCl | 298 | 100 | 1.0551 |
| 10 | $T_{298}C_6$ | 6 M NaCl | 298 | 100 | 0.7055 |
| 11 | $T_{310}C_0$ | Pure Water | 310 | 100 | 2.8333 |
| 12 | $T_{310}C_1$ | 1 M NaCl | 310 | 100 | 2.5831 |
| 13 | $T_{310}C_2$ | 2 M NaCl | 310 | 100 | 2.1387 |
| 14 | $T_{310}C_4$ | 4 M NaCl | 310 | 100 | 1.3777 |
| 15 | $T_{310}C_6$ | 6 M NaCl | 310 | 100 | 0.9773 |
| 16 | $T_{310}C_0P_{200}$ | Pure Water | 310 | 200 | 2.7107 |
| 17 | $T_{310}C_1P_{200}$ | 1 M NaCl | 310 | 200 | 2.5553 |
| 18 | $T_{310}C_2P_{200}$ | 2 M NaCl | 310 | 200 | 2.1327 |
| 19 | $T_{310}C_4P_{200}$ | 4 M NaCl | 310 | 200 | 1.4887 |
| 20 | $T_{310}C_6P_{200}$ | 6 M NaCl | 310 | 200 | 0.9889 |
| 21 | $T_{310}C_0P_{300}$ | Pure Water | 310 | 300 | 2.8552 |
| 22 | $T_{310}C_1P_{300}$ | 1 M NaCl | 310 | 300 | 2.3110 |
| 23 | $T_{310}C_2P_{300}$ | 2 M NaCl | 310 | 300 | 2.0237 |
| 24 | $T_{310}C_4P_{300}$ | 4 M NaCl | 310 | 300 | 1.4107 |
| 25 | $T_{310}C_6P_{300}$ | 6 M NaCl | 310 | 300 | 0.8889 |
| 26 | $T_{323}C_0$ | Pure Water | 323 | 100 | 3.8327 |
| 27 | $T_{323}C_1$ | 1 M NaCl | 323 | 100 | 3.1553 |
| 28 | $T_{323}C_2$ | 2 M NaCl | 323 | 100 | 2.4940 |
| 29 | $T_{323}C_4$ | 4 M NaCl | 323 | 100 | 1.8331 |
| 30 | $T_{323}C_6$ | 6 M NaCl | 323 | 100 | 1.3443 |
| 31 | $T_{348}C_0$ | Pure Water | 348 | 100 | 5.6167 |
| 32 | $T_{348}C_1$ | 1 M NaCl | 348 | 100 | 4.6385 |
| 33 | $T_{348}C_2$ | 2 M NaCl | 348 | 100 | 3.9663 |
| 34 | $T_{348}C_4$ | 4 M NaCl | 348 | 100 | 2.9831 |
| 35 | $T_{348}C_6$ | 6 M NaCl | 348 | 100 | 1.9271 |
| 36 | $T_{348}C_0P_{200}$ | Pure Water | 348 | 200 | 5.4443 |
| 37 | $T_{348}C_1P_{200}$ | 1 M NaCl | 348 | 200 | 4.1440 |
| 38 | $T_{348}C_2P_{200}$ | 2 M NaCl | 348 | 200 | 3.9110 |
| 39 | $T_{348}C_4P_{200}$ | 4 M NaCl | 348 | 200 | 2.6553 |
| 40 | $T_{348}C_6P_{200}$ | 6 M NaCl | 348 | 200 | 1.9999 |
| 41 | $T_{348}C_0P_{300}$ | Pure Water | 348 | 300 | 5.2110 |
| 42 | $T_{348}C_1P_{300}$ | 1 M NaCl | 348 | 300 | 4.4777 |
| 43 | $T_{348}C_2P_{300}$ | 2 M NaCl | 348 | 300 | 4.0556 |
| 44 | $T_{348}C_4P_{300}$ | 4 M NaCl | 348 | 300 | 2.6999 |
| 45 | $T_{348}C_6P_{300}$ | 6 M NaCl | 348 | 300 | 1.7444 |
| 46 | $T_{373}C_0$ | Pure Water | 373 | 100 | 7.0820 |
| 47 | $T_{373}C_1$ | 1 M NaCl | 373 | 100 | 5.7219 |
| 48 | $T_{373}C_2$ | 2 M NaCl | 373 | 100 | 5.2610 |
| 49 | $T_{373}C_4$ | 4 M NaCl | 373 | 100 | 3.9883 |
| 50 | $T_{373}C_6$ | 6 M NaCl | 373 | 100 | 2.8830 |
| 51 | $T_{398}C_0$ | Pure Water | 398 | 100 | 9.7333 |
| 52 | $T_{398}C_1$ | 1 M NaCl | 398 | 100 | 8.4443 |
| 53 | $T_{398}C_2$ | 2 M NaCl | 398 | 100 | 6.3440 |
| 54 | $T_{398}C_4$ | 4 M NaCl | 398 | 100 | 5.3663 |
| 55 | $T_{398}C_6$ | 6 M NaCl | 398 | 100 | 3.8887 |
| 56 | $T_{423}C_0$ | Pure Water | 423 | 100 | 10.8773 |
| 57 | $T_{423}C_1$ | 1 M NaCl | 423 | 100 | 9.5553 |
| 58 | $T_{423}C_2$ | 2 M NaCl | 423 | 100 | 8.1333 |
| 59 | $T_{423}C_4$ | 4 M NaCl | 423 | 100 | 5.9997 |
| 60 | $T_{423}C_6$ | 6 M NaCl | 423 | 100 | 4.7497 |
| 61 | $T_{423}C_0P_{200}$ | Pure Water | 423 | 200 | 11.2886 |
| 62 | $T_{423}C_1P_{200}$ | 1 M NaCl | 423 | 200 | 8.7773 |
| 63 | $T_{423}C_2P_{200}$ | 2 M NaCl | 423 | 200 | 8.4443 |
| 64 | $T_{423}C_4P_{200}$ | 4 M NaCl | 423 | 200 | 5.7775 |
| 65 | $T_{423}C_6P_{200}$ | 6 M NaCl | 423 | 200 | 4.9111 |
| 66 | $T_{423}C_0P_{300}$ | Pure Water | 423 | 300 | 11.2775 |
| 67 | $T_{423}C_1P_{300}$ | 1 M NaCl | 423 | 300 | 9.0553 |
| 68 | $T_{423}C_2P_{300}$ | 2 M NaCl | 423 | 300 | 7.3442 |
| 69 | $T_{423}C_4P_{300}$ | 4 M NaCl | 423 | 300 | 6.1109 |
| 70 | $T_{423}C_6P_{300}$ | 6 M NaCl | 423 | 300 | 4.9889 |

To explore the reasons behind the behavior of the $CO_2$ diffusion coefficient in NaCl brine, we performed the RDF and H-bond analyses in each case separately. The RDF for Na-$O_{H_2O}$, Na-$O_{CO_2}$, Na-Cl and $O_{CO_2}$-$H_{H_2O}$ in two temperatures of 294 K and 423 K are shown in **Figure *3*** to **6**, respectively. The number of H-bonds between $O_{CO_2}$-$H_{H_2O}$ and $O_{H_2O}$-$H_{H_2O}$ are also illustrated in **Table 2**. First, we scrutinized the RDF figures and H-bond values to determine the impacts of the aforementioned parameters on intermolecular interactions between species. Then, we concluded the reason behinds the variation in $CO_2$ diffusion coefficient under different circumstances.

According to **Figure 3**, in general description, an increment in temperature reduces the magnitude of the first peak at each constant NaCl salinity concentration. Also, as the NaCl salinity rises, the height of the first peaks increases at relatively low temperatures, while at high temperatures (e.g., 423 K) go through a declining trend instead. In all cases, regardless of the salinity and temperature, the first hydration shell of $Na^+$, identified by the first peak location of Na-$O_{H_2O}$, appears at 2.4 Å, in line with previous studies [32, 62-64]. The second hydration shell of $Na^+$ is formed at approximately 4.5 Å, in which the height of the RDFs has a reverse order compared to the first peak values. It is worth noting that the second peak location of Na-$O_{H_2O}$ has a negligible dependency on the temperature variation. The impacts of these changes can be explained by the dependency of intermolecular interactions of $Na^+$ and water molecules on the concentration and temperature. Regarding the impact of concentration, it is clear that the presence of more cations enhances the possibility of interactions with water molecules, resulting in more hydration shells. That is why a higher height is observed for more saline water. In respect of the temperature parameter, $Na^+$ ions are hydrated by water molecules, by which a strong attractive force is formed between water molecules and the surrounded ions. The temperature increase leads to $Na^+$ hydration shell size reduction, which triggers water molecules to decrease around the cations, facilitating the movement of $Na^+$ cations. In addition, the movement of water molecules increases because a rise in temperature intensifies repulsive forces among water molecules. This behavior that is shown by reduction in the height of RDFs with temperature increment is also confirmed by comparing the number of H-bond between water molecules provided in **Table *2***, which offers a decrease of intermolecular interactions among water molecules due to temperature increment. Nevertheless, in the case of the exception (T= 423 K), there is no intense interaction between $Na^+$ and water molecules due to the higher NaCl salinity. In fact, the heating effect of the temperature may trigger spreading cations owing to a robust repulsive force among them, leading to a decrease in the possibility of the interactions with water molecules. That is why the height of the RDF is lower for the higher concentration at 423 K. Note that the RDFs of all cases are shown in **Figure S2**.

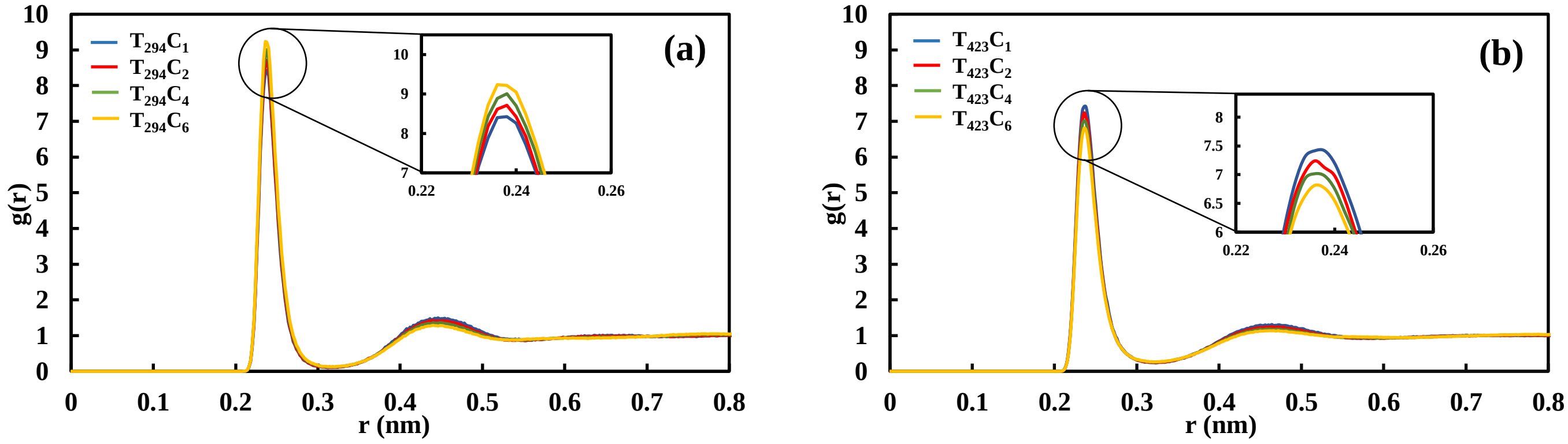

**Figure 3**- RDF curves of $Na\text{-}O_{H_2O}$ in various NaCl solutions at the temperatures of (a) 294 K, and (b) 423 K

**Figure *4*-** demonstrates the RDF values of Na-$O_{CO_2}$ for two cases of 294 K and 423 K under various NaCl concentrations. RDFs of Na-$O_{CO_2}$ for the rest of cases are shown in **Figure S3**. As is shown, the RDF values experience a declining trend with temperature increment, in which the temperature changes have a relatively severe impact on the height of the RDFs, attributing to the higher NaCl concentrations.

Nevertheless, the distances between $Na^+$ and $O_{CO_2}$, identified by the first and second peak location of RDF, are fairly independent of temperature changes by taking the average distance of 2.58 Å and 5 Å, respectively. On the other hand, considering the salinity variation, the larger height of RDFs belongs to the higher NaCl concentrations at a constant temperature. Compared to the intermolecular interactions of $Na^+$ and water molecules, for all cases, rising the temperature decreases the affinity of $Na^+$ ions towards $O_{CO_2}$ for each concentration, whereas the salinity increment enhances the intermolecular interplays between the cations and $CO_2$ molecules.

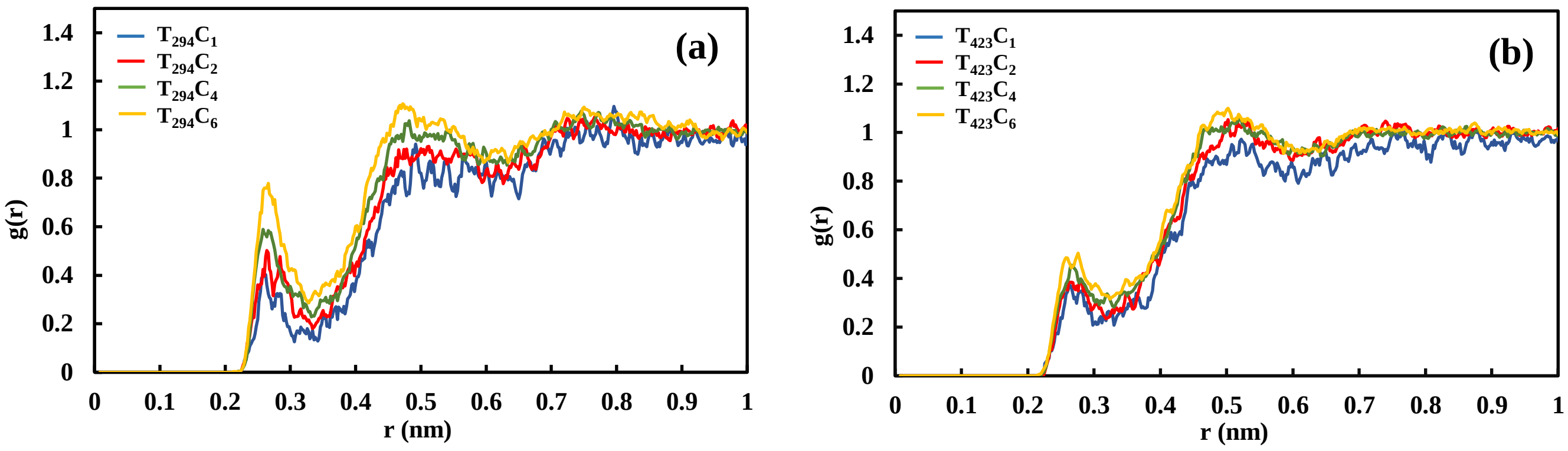

**Figure 4**-RDF curves of $Na\text{-}O_{CO_2}$ in various NaCl solutions at the temperatures of (a) 294 K, and (b) 423 K.

RDF curves for Na-Cl at temperatures of 294 K and 423 K are demonstrated in ***Figure 5***. As is evident, although increasing temperature reduces the interaction intensity of Na-$O_{H_2O}$ and Na-$O_{CO_2}$, the intermolecular interactions of Na-Cl considerably enhance instead. Furthermore, the higher NaCl salinity concentration forms stronger interaction between Na-Cl. It can be inferred that at low temperatures, $Na^+$ cations are mainly hydrated by water molecules, and some of them lean towards $CO_2$. However, the temperature increment reduces the interplays of water and $CO_2$ with $Na^+$ cations. Indeed, more interactions are created between Na-Cl (see **Figure S4** for the rest of the cases).

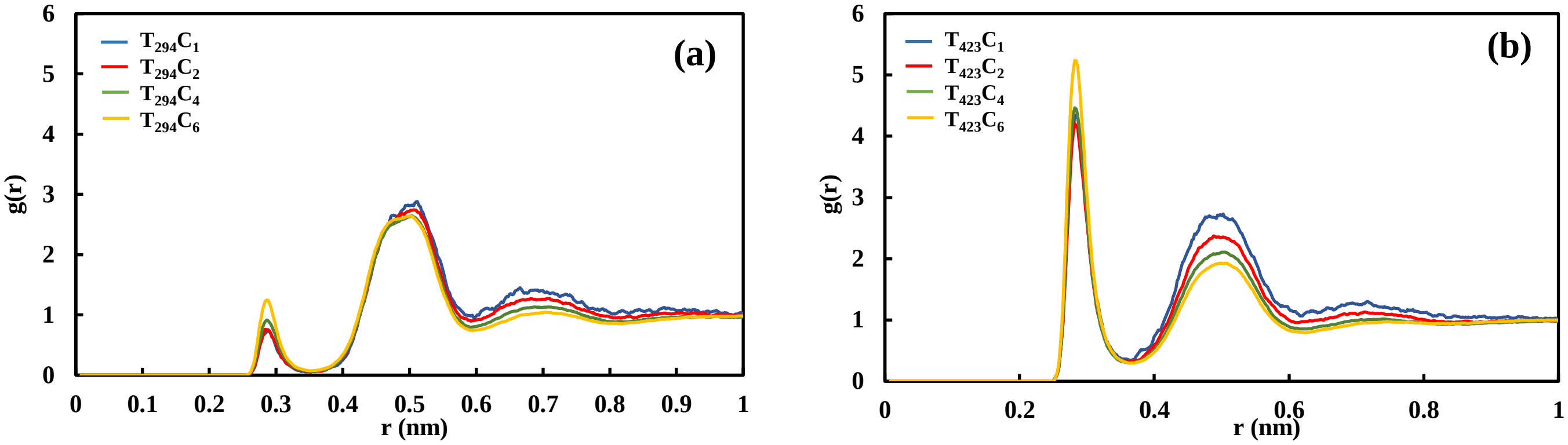

**Figure 5**- RDF curves of $Na$-$Cl$ in various NaCl solutions at the temperatures of a) 294 K, and b) 423 K

The RDFs for the pair of $O_{CO_2}$and $H_{H_2O}$ in temperature of 294 K and 423 K are presented in **Figure 6-** RDF curves of $O_{CO_2}$-$H_{H_2O}$ in various aqueous solutions at the temperatures of (a) 294 K, and (b) 423 K.

(see **Figure S4** for the rest of the temperatures). It seems that an increase in temperature enhances the affinity of $H_{H_2O}$ towards the $O_{CO_2}$since the slight jump in the first peak of RDFs at an average distance of 2 Å, which are linked to the presence of the H-bonds. The ascending trend of the number of H-bond between $CO_2$ and water molecules versus temperature increment proves the results of RDFs. According to **Table 2**, for the 1 M NaCl salinity concentration as an example, the number of H-bond between $CO_2$ and water molecules is 4.4352 and 6.1119 at 294 K and 423 K, respectively, indicating the stronger repulsive force among water molecules at elevated temperatures, which lead to improving the possibility of $CO_2$ surrounding by water molecules due to the temperature increase. Nevertheless, the second peak location of RDFs shifted to 3.9 Å from 3.4 Å at temperatures of 294 K and 423 K, respectively, identifying the weakening of the intermolecular interactions of the pairs at the larger distance. Furthermore, the presence of more $Na^+$ ions reinforces interactions of $O_{CO_2}$-$H_{H_2O}$ pairs. In more detail, when NaCl salinity concentration increases, more ionic hydration shells are also created. On the other hand, the solid electrostatic interactions between cations in the hydrated form with $CO_2$ leads to locating more water molecules close to the $CO_2$. Subsequently, the intermolecular interplays between $CO_2$ and water molecules increase. **Table 2** also reveals the higher number of H-bond between $O_{CO_2}$-$H_{H_2O}$ pairs at more saline brine for all considered temperatures.

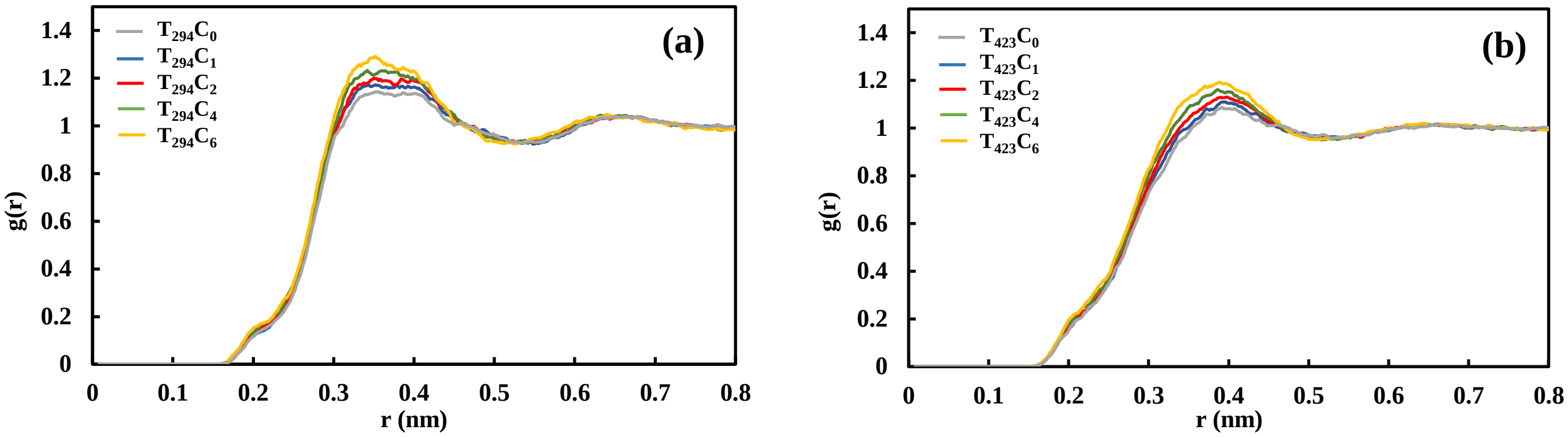

**Figure 6**- RDF curves of $O_{CO_2}$-$H_{H_2O}$ in various aqueous solutions at the temperatures of (a) 294 K, and (b) 423 K.

**Table 2**-The H-bonds values for $O_{CO_2}$-$H_{H_2O}$ and $O_{H_2O}$-$H_{H_2O}$.

| | $O_{CO_2} - H_{H_2O}$ | | | | $O_{H_2O} - H_{H_2O}$ | | | |
|---|---|---|---|---|---|---|---|---|
| Temperature (K) \ Molarity (M) | 1 | 2 | 4 | 6 | 1 | 2 | 4 | 6 |
| 294 | 4.4352 | 4.6290 | 4.7613 | 4.8189 | 4798.8 | 4296.8 | 3380.3 | 2677.5 |
| 298 | 4.0551 | 4.6623 | 4.7672 | 4.9011 | 4730.7 | 4292.7 | 3386.8 | 2661.5 |
| 310 | 4.7107 | 4.8112 | 5.0119 | 5.0779 | 4687.4 | 4218.6 | 3330.1 | 2616.9 |
| 323 | 5.0839 | 5.2061 | 5.2235 | 5.2453 | 4613.6 | 4126.1 | 3306.2 | 2597.8 |
| 348 | 5.4535 | 5.4875 | 5.6154 | 5.7077 | 4498.5 | 3987.4 | 3191.7 | 2558.8 |
| 373 | 5.8182 | 5.9191 | 6.0083 | 6.0684 | 4308.9 | 3897.9 | 3089.6 | 2460.9 |
| 398 | 5.9445 | 6.0259 | 6.1114 | 6.2649 | 4145.9 | 3703.6 | 3030.1 | 2455.9 |
| 423 | 6.1119 | 6.1334 | 6.1909 | 6.2816 | 4001.8 | 3581.4 | 2952.3 | 2372.9 |

Overall, the increasing NaCl salinity not only enhances the affinity of $Na^+$ cations towards $CO_2$ molecules to form direct interactions but also locates more and more water molecules around themselves to form hydration shells. The formation of direct interactions and hydration shells impede the movement of $CO_2$ molecules. That is why the higher NaCl salinity restricts the freedom of $CO_2$ molecules, causing a decrease in the diffusion coefficient. Regarding the heat effect of the temperature, the rise in the $CO_2$ diffusion at elevated temperature originates from intensifying repulsive forces among $Na^+$ cations, resulting in lower intermolecular interaction with $CO_2$ molecules. Furthermore, since the hydration shells serve as a $CO_2$ movement barrier, a temperature rise reduces the size of hydration shells, providing more freedom for $CO_2$ molecules. There are the reasons behind higher $CO_2$ diffusion at elevated temperatures. It is worth noting that although $CO_2$ molecules are more surrounded by water molecules separated from the cation's hydration shells due to the temperature increment, their interaction is not strong enough to restrict the freedom of $CO_2$ molecules as a result of lower $CO_2$ interactions with $Na^+$ cations and decrease in size of hydration shells.

## 3.2. Pressure Effect on the $CO_2$ Diffusion Coefficient

Our investigation includes further analysis of pressure effects on the $CO_2$ diffusion coefficient, enabling us to obtain an in-depth understanding of critical parameters. According to the majority of previous studies, pressure does not significantly affect the diffusion coefficient, at least to a specific temperature and pressure [28, 65]. Nevertheless, Zhang et al. [16] claimed that the diffusion coefficient increases linearly with pressure, but this rise becomes smaller and smaller at some point. The results of Moultos et al. [36] demonstrated that the pressure effect on diffusion coefficient becomes more pronounced with increasing temperature, whereas before that, the change is negligible. In this study, simulations were repeated at 20 and 30 MPa at three levels of temperature: low (323 K), mediocre (373 K), and high (423 K). It was found that no substantial changes are present in the $CO_2$ diffusion coefficient under different pressures. According to **Figure *7***, it can be seen that there is neither a significant change in diffusion coefficients at various pressures nor does a remarkable trend indicate the coefficients are leaning toward higher or lower values with increasing pressure.

We also examined RDF values to corroborate our diffusion coefficient findings. These results can be seen in **Figures S6-S12**. In general, neither the peak values nor their distances change significantly at various pressures. For instance, the first and second peaks of Na-$O_{H_2O}$ at 313 K and 10 MPa appear at 2.4 and 4.5 Å, respectively, and the same pattern is observed at 20 and 30 MPa, as well. Hence, the similarity in RDFs values of various pressures is aligned with the fact that pressure has no noticeable impact on the $CO_2$ diffusion coefficient.

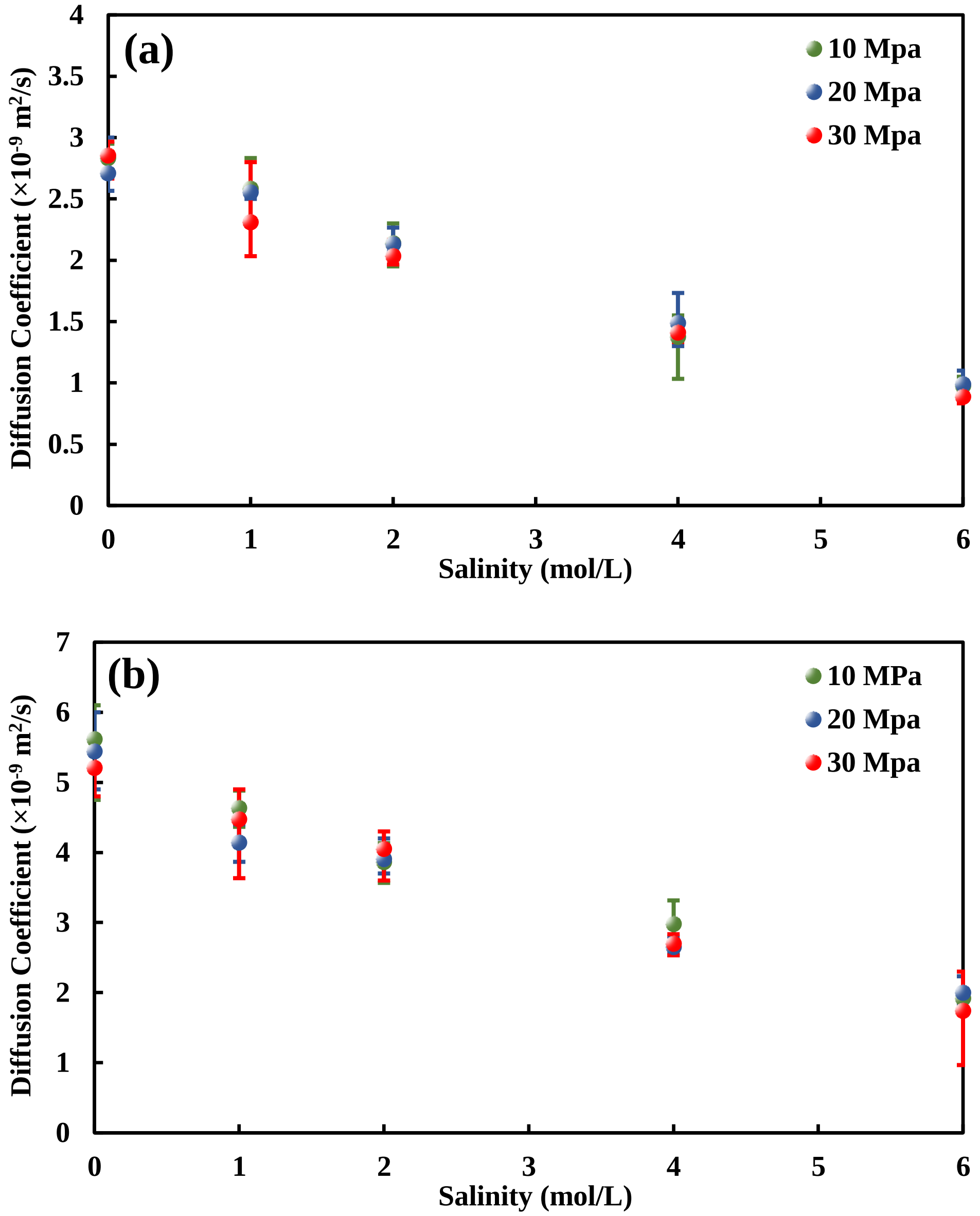

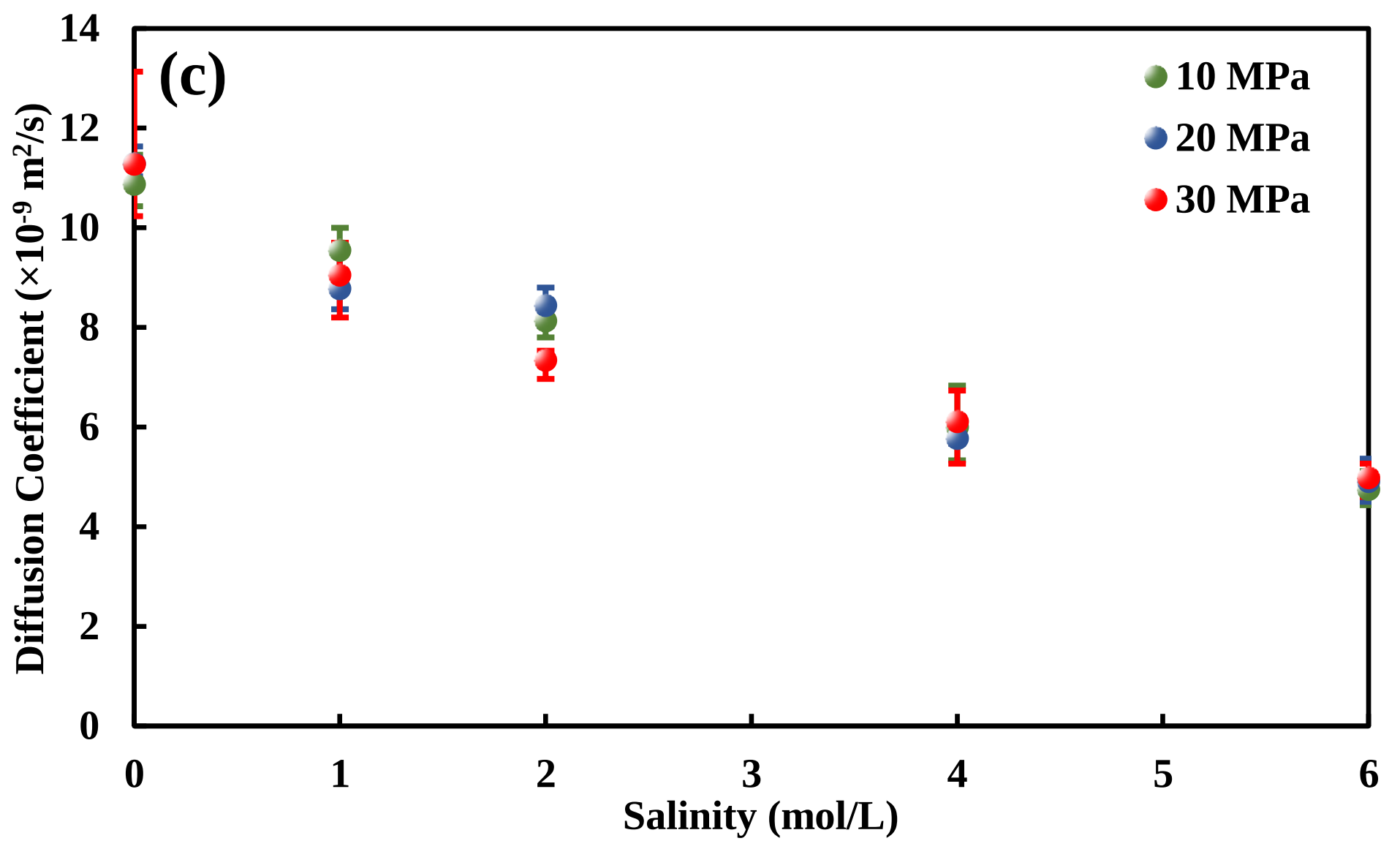

**Figure 7**- The $CO_2$ diffusion coefficient in 0-6 M NaCl solution under three different pressures of 10, 20, and 30 MPa and temperatures of (a) 310 K, (b) 348 K, and (c) 423 K.

## 3.3. Type of Salt Effect on the $CO_2$ Diffusion Coefficient

As is known, NaCl is the most common salt in aquifers, but other salt compositions may also be present. In fact, studies on the $CO_2$ diffusion coefficient in brine under the influence of various salt compositions have been limited to determining if any difference exists among them. Zhang et al. [16] considered four types of salts other than NaCl at 298 K and reported that the diffusion coefficient of $CO_2$ has the lowest and highest value in the presence of $NaHCO_3$ and $CaCl_2$, respectively. In 2015, Cadogan et al. [29] studied the $CO_2$ diffusion coefficients in brine containing NaCl, $CaCl_2$, and $Na_2SO_4$ separately. They found that NaCl has the smallest impact on the $CO_2$ diffusion coefficient compared to pure water, while $Na_2SO_4$ solution has the major contribution in changing the $CO_2$ diffusion coefficient.

**Figure 8** displays the values of the $CO_2$ diffusion coefficient under the effect of different salts, including $MgCl_2$, $CaCl_2$, $Na_2SO_4$, and KCl. The simulations were carried out at 323 K and two levels of concentration (1 and 2 M). According to the results, $CaCl_2$ and KCl have the most and the least impact on the $CO_2$ diffusion coefficients compared to pure water, respectively. This indicates that analyzing water compositions regarding existent salts and their concentrations should be considered for the CCS process, particularly the $CO_2$ diffusion coefficient. Furthermore, the effect of different salts at higher concentrations on the $CO_2$ dissolution behavior is more prominent. **Table *3*** lists the results of this section.

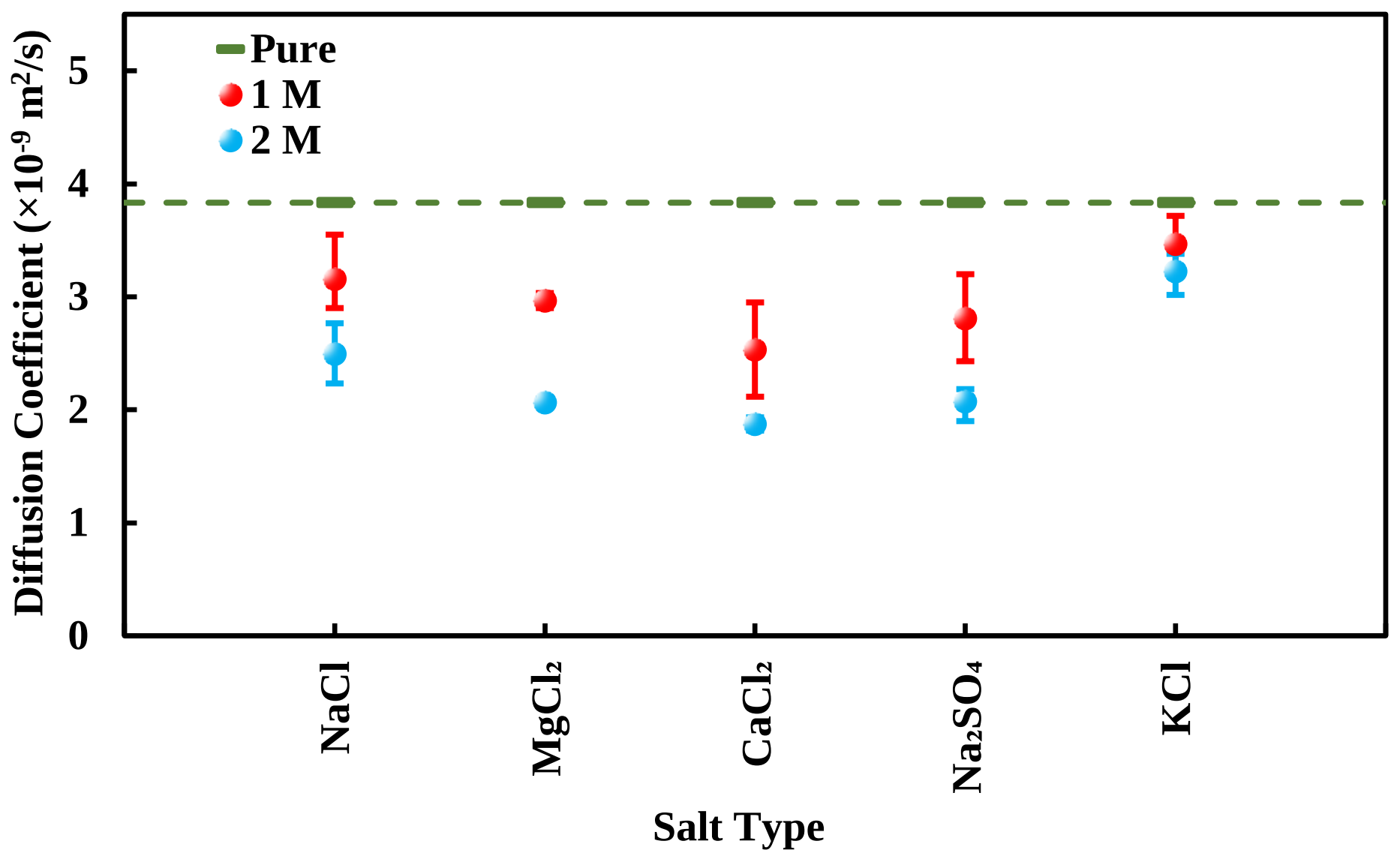

**Figure 8**- The $CO_2$ diffusion coefficient in brine in the presence of different salts under temperature of 323 K and pressure of 10 MPa.

**Table 3**- The results of $CO_2$ diffusion coefficient in various brine solutions.

| No. | Case Name | Solution $\left(\frac{mol}{L}\right)$ | Temperature (K) | Pressure (MPa) | D $(\times 10^{-9}\,\frac{m^2}{s})$ |
|---|---|---|---|---|---|
| 1 | $T_{323}C_1S_{MgCl_2}$ | 1 M $MgCl_2$ | 323 | 100 | 2.9665 |
| 2 | $T_{323}C_2S_{MgCl_2}$ | 2 M $MgCl_2$ | 323 | 100 | 2.0663 |
| 3 | $T_{323}C_1S_{CaCl_2}$ | 1 M $CaCl_2$ | 323 | 100 | 2.5330 |
| 4 | $T_{323}C_2S_{CaCl_2}$ | 2 M $CaCl_2$ | 323 | 100 | 1.8747 |
| 5 | $T_{323}C_1S_{Na_2SO_4}$ | 1 M $Na_2SO_4$ | 323 | 100 | 2.8100 |
| 6 | $T_{323}C_2S_{Na_2SO_4}$ | 2 M $Na_2SO_4$ | 323 | 100 | 2.0721 |
| 7 | $T_{323}C_1S_{KCl}$ | 1 M KCl | 323 | 100 | 3.4660 |
| 8 | $T_{323}C_2S_{KCl}$ | 2 M KCl | 323 | 100 | 3.2266 |

We analyzed the RDF values for the involved components to obtain the relationship between the salt composition and the $CO_2$ diffusion coefficient. Generally, it can be said that due to the difference in the nature of salt components, their intermolecular interactions with water and $CO_2$ are varied. That is why no straightforward relationship is defined for the salt compositions and the $CO_2$ diffusion coefficient. Nevertheless, there are some valuable points regarding the mechanisms of salts on changing the $CO_2$ diffusion that should be discussed. **Figure 9a** shows that the first hydration shell of $Na^+$, $Mg^{2+}$, $Ca^{2+}$, and $K^+$ are located at 2.4, 2.1, 2.4, and 2.8 Å, respectively, which is consistent with previous studies [45, 66]. As is clear, the maximum intensity of the cation interaction with water molecules belongs to $Mg^{2+}$, $Ca^{2+}$, $Na^+$ of NaCl, $K^+$, and $Na^+$ of $Na_2SO_4$, respectively. While the higher tendency of the cation interaction with water molecules, the less intensity of the cation interaction with $CO_2$ will result, except $Na^+$ of $Na_2SO_4$ (see **Figure 9b** ). On the other hand, comparing the RDFs of cations with anions reveals that most cations in $Na_2SO_4$ solutions tend to interact with anions rather than water or $CO_2$ molecules, as shown in **Figure 10a**. In other words, the strong interaction between $Na^+$ and $SO_4^{2-}$ explains the low hydration tendency of $Na^+$ ions and their less affinity towards $CO_2$ molecules. The intense interaction between cations and anions is also evident in $MgCl_2$. This also can justify the low interaction of $Mg^{2+}$ with $CO_2$ molecules. **Figure 10b** illustrates RDFs of $H_{H_2O}$-$O_{H_2O}$ that have the same trends for all cases. However, in the case of $Na_2SO_4$, since ionic compositions have a higher tendency towards each other, water molecules also have more freedom to interact with $CO_2$. According to the intermolecular interactions of cations, it can be concluded that salts can reduce the movement of $CO_2$ molecules in saline brine following three possible mechanisms of the direct interaction of cations with $CO_2$, formation of actions' hydration, and aggregation of ionic components. In further detail, monovalent cations with low hydration enthalpy, which have no robust tendency towards water, can interplay with $CO_2$ molecules directly, leading to the interruption of $CO_2$ freedom (e.g., KCl). Cations with a strong attraction towards the water molecules can form hydration shells, acting as a barrier for the movement of $CO_2$ (e.g., $MgCl_2$). In addition, ionic compositions with the potential of aggregation formation (e.g., $Na_2SO_4$) can restrict the $CO_2$ diffusion in saline brine. Also, the formed ionic cluster occupies a large amount of space, causing less volume available for $CO_2$ molecules. Note that brine like $MgCl_2$ may influence the $CO_2$ diffusion coefficient by following two mechanisms. **Figure *11*** illustrates the final snapshots of the simulation box containing different salts.

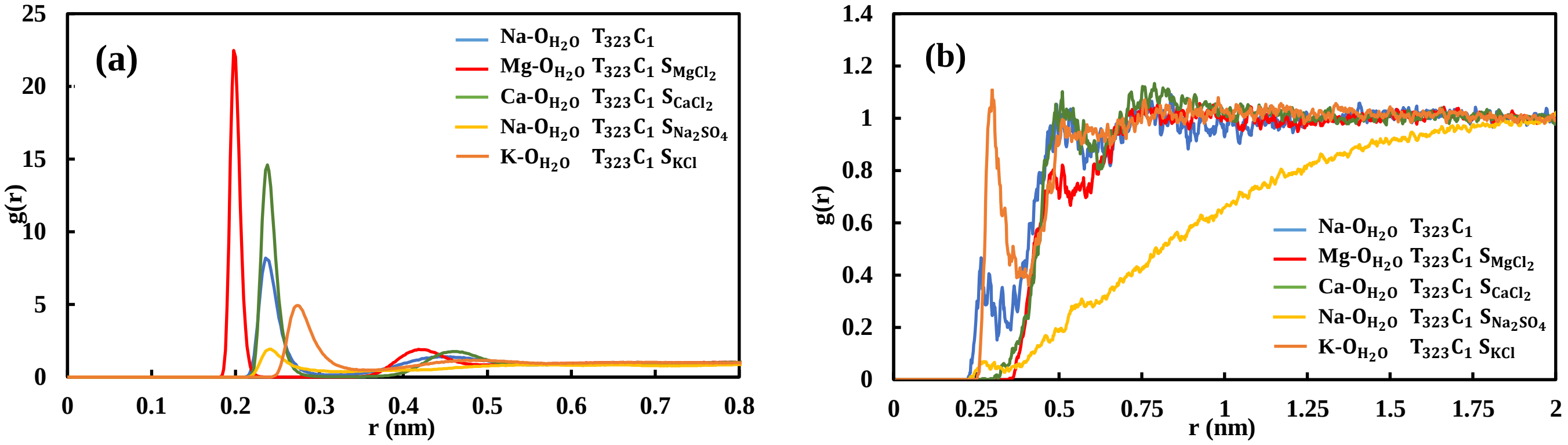

**Figure 9**- The RDF values between cations and (a) $O_{H_2O}$ and (b) $O_{CO_2}$ in various saline solutions at the temperatures 323 K and 1 M concentration.

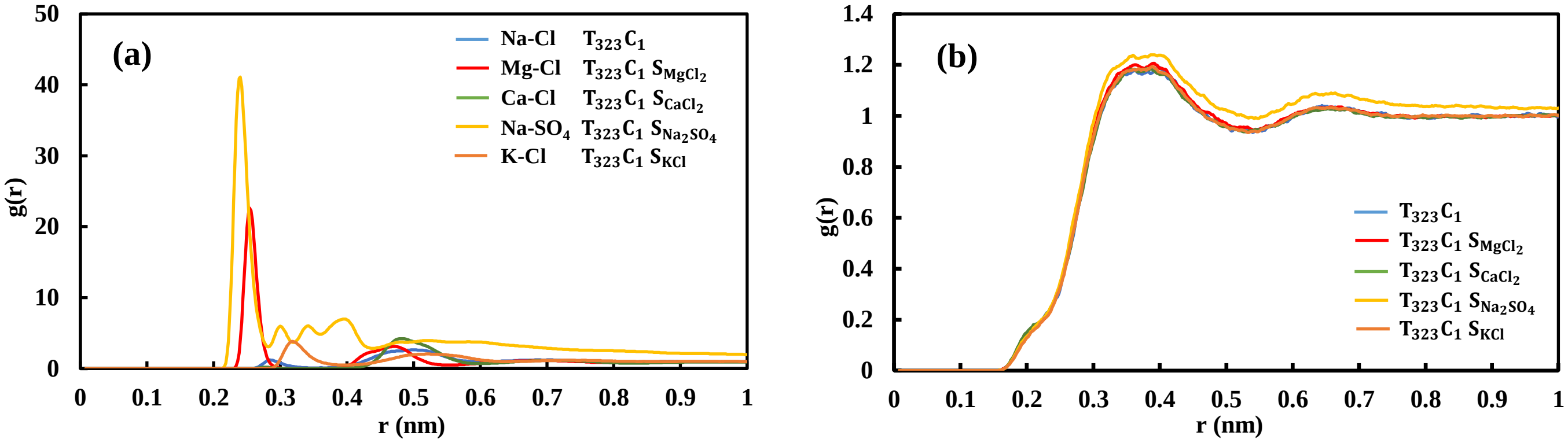

**Figure 10**- The RDF values between (a) cations-anions and (b) $H_{H_2O}$-$O_{H_2O}$ in various saline solutions at the temperatures 323 K and 1 M concentration.

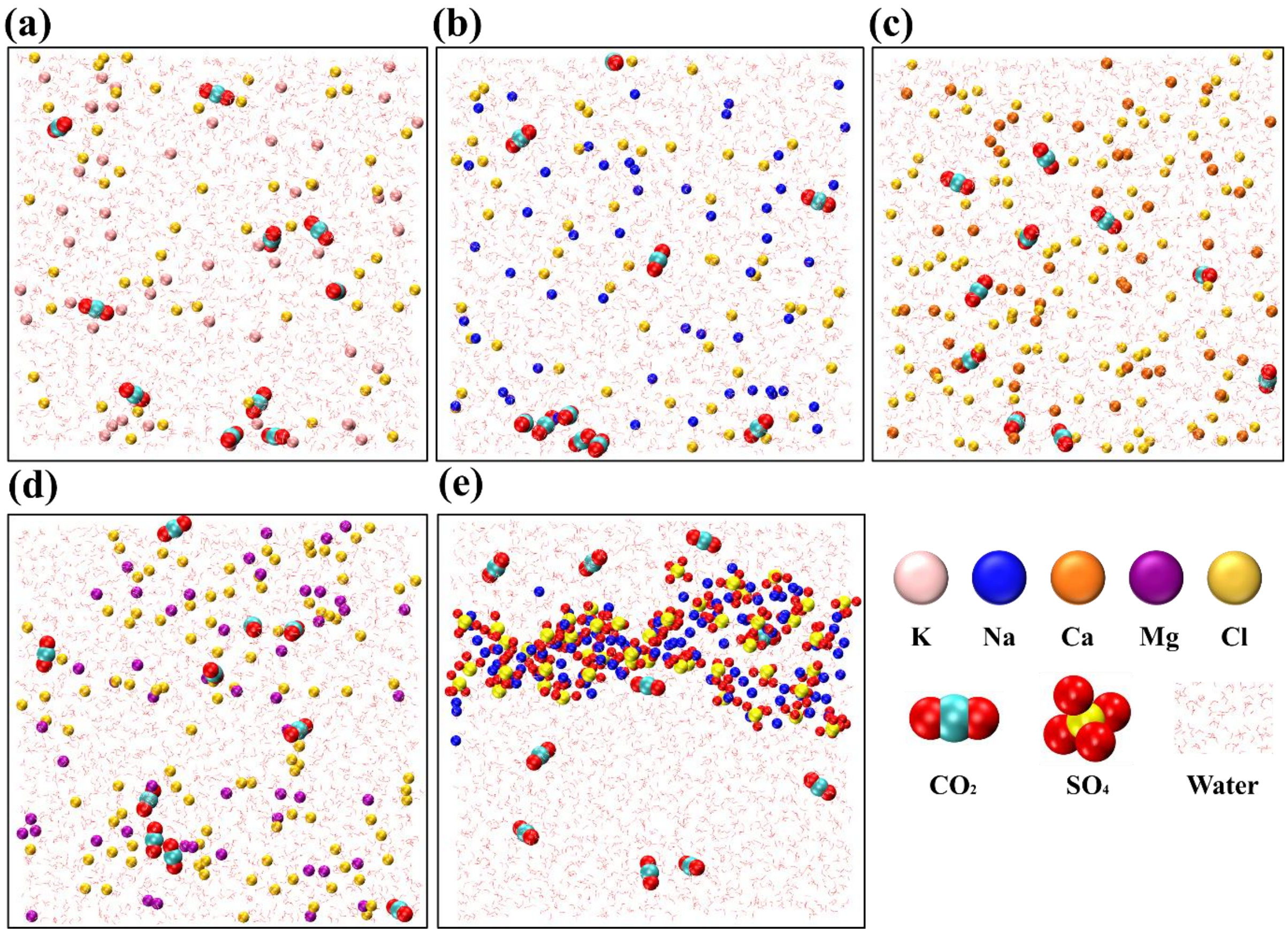

**Figure 11-** Final Snapshots of simulation systems containing solution with the name of (a) $T_{323}C_1S_{KCl}$, (b) $T_{323}C_1S_{NaCl}$, (c) $T_{323}C_1S_{CaCl_2}$, (d) $T_{323}C_1S_{MgCl_2}$, and (e) $T_{323}C_1S_{Na_2SO_4}$. As is clear, the tendency of ionic compositions to create aggregations is in order of (e)> (d)> (c)> (b)> (a).

## 3.4. $CO_2$ Dissolution Behavior

In order to get further insight into the difference between considering and ignoring the change of $CO_2$ diffusion coefficient in $CO_2$ dissolution behavior, we considered the cases of $CO_2$ diffusion in NaCl brine at 323 K for direct numerical simulation. **Figure *12*** illustrates the averaged dissolved $CO_2$ (normalized by maximum solubility) and $CO_2$ dissolution flux for each case (dashed curves are those without considering the salt effect, and their names include * at the end). The lack of consideration of diffusion coefficient changes in 1M NaCl solution results in a wrong evaluation of dissolved $CO_2$ and an overestimation of dissolution flux. Furthermore, without considering the actual $CO_2$ diffusion coefficient at higher salinities, both dissolved $CO_2$ and dissolution flux would seem to be excessively high at higher salinities (e.g., 4M). Considering the initial stage of the $CO_2$ dissolution process (beginning of the curves of dissolved $CO_2$ in **Figure *12*a**), all cases based on the pure water $CO_2$ diffusion coefficient show a higher amount of dissolved $CO_2$, suggesting the overestimation of the diffusion mechanism strength. It can be inferred that predicting $CO_2$ dissolution process parameters without considering the correct diffusion coefficient results in massive errors. **Figure *13*** displays the dissolved $CO_2$ pattern for cases $T_{323}C_0$, $T_{323}C_1$, $T_{323}C_1^*$, $T_{323}C_4$, and $T_{323}C_4^*$. Even if we disregard the differences in dissolved $CO_2$ and dissolution flux, time differences can significantly impact the long-term $CO_2$ sequestration planning. In addition, these behaviors are more pronounced at higher temperatures. Considering the same salinity, if pure water $CO_2$ diffusion is used instead of the real brine diffusion coefficient, then the damping capability of the system for gravitational instabilities increases unrealistically. Therefore, the relative importance of the convective to diffusive mechanism decreases, and a delay in dissolution regimes changes is observed. It should be noted that the transition times between dissolution regimes is controlled by not only diffusion coefficient but also density difference, which is a critical factor in the solubility process. That is why cases with 4 M salinity show the latest transition times. Since the solubility values are the lowest in these cases, the density difference, considered the driving force for the convective flow, would be the lowest. At all, in spite of the minimum diffusion coefficient for assessed cases, the density difference has more control on the system, and it increases the transition times.

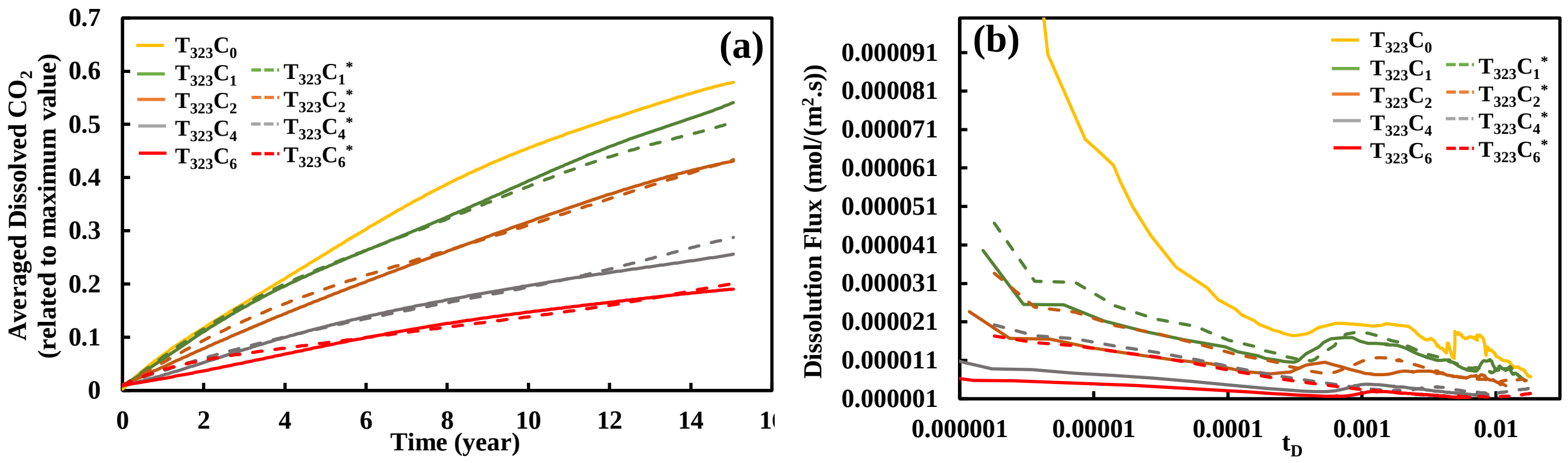

**Figure 12**- (a) Averaged dissolved $CO_2$ and (b) dissolution flux for cases at 323 K.

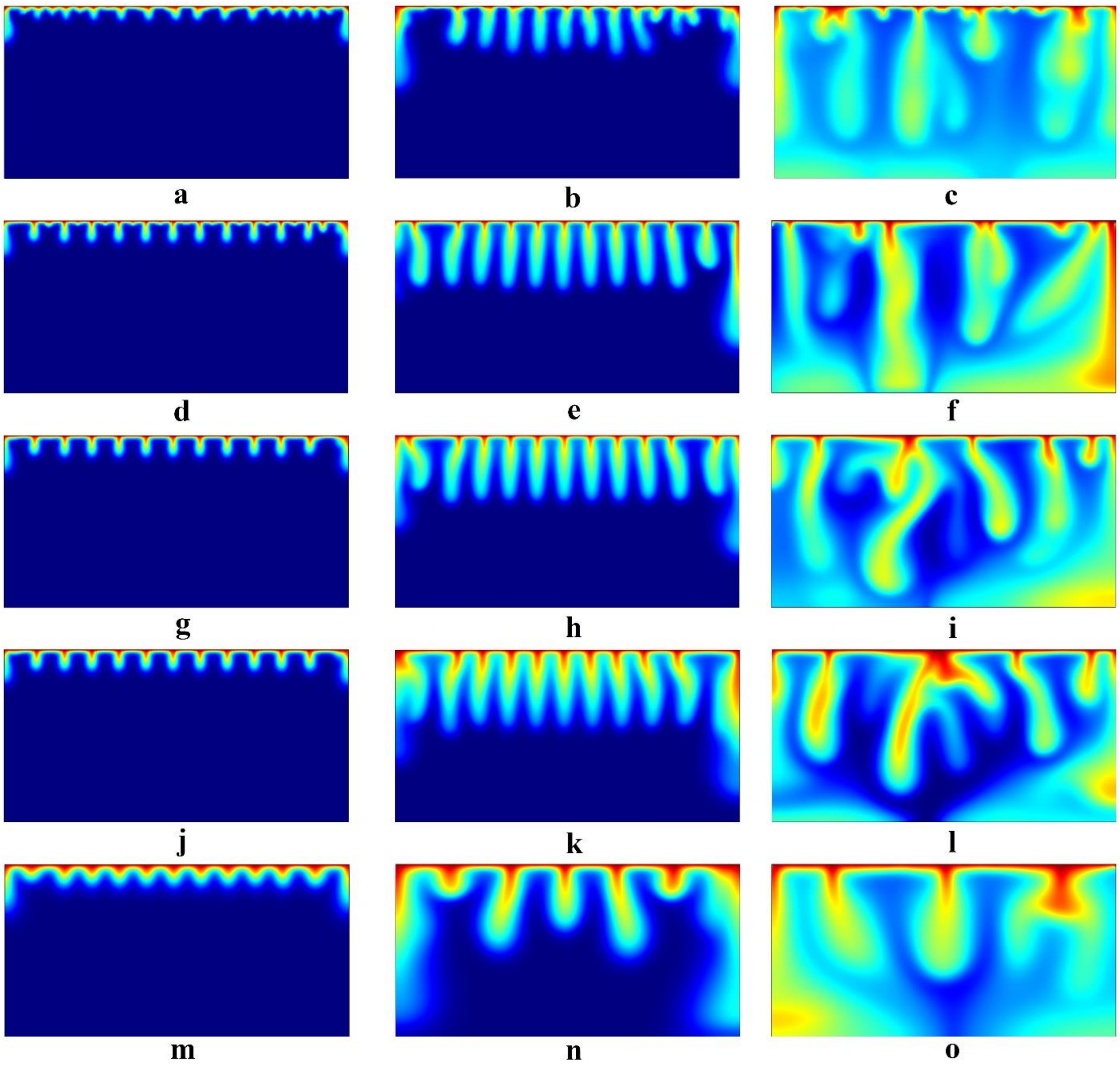

**Figure 13-** Dissolved $CO_2$ patterns for cases $T_{323}C_0$ (first row), $T_{323}C_1$ (second row), $T_{323}C_1^*$ (third row), $T_{323}C_4$ (fourth row), and $T_{323}C_4^*$ (fifth row): a) $T_{323}C_0$ at onset of convection (0.25 year), b) $T_{323}C_0$ in the quasi-steady state (0.4 years), c) $T_{323}C_0$ at the onset of shut-down (7.5 years), d) $T_{323}C_1$ at onset of convection (0.31 year), e) $T_{323}C_1$ in the quasi-steady state (0.6 years), f) $T_{323}C_1$ at the onset of shut-down (9 years), g) $T_{323}C_1^*$ at onset of convection (0.35 years), h) $T_{323}C_1^*$ in the quasi-steady state (0.65 years), i) $T_{323}C_1^*$ at the onset of shut-down (10.25 years), j) $T_{323}C_4$ at onset of convection (0.8 year), k) $T_{323}C_4$ in the quasi-steady state (1.6 years), l) $T_{323}C_4$ at the onset of shut-down(20.4 years), m) $T_{323}C_4^*$ at onset of convection (1.4 years), n) $T_{323}C_4^*$ in the quasi-steady state (2.4 years), o) $T_{323}C_4^*$ at the onset of shut-down (21.8 years).

# 4. Conclusions

In this research, a comprehensive study on the $CO_2$ diffusion coefficient in water/brine under a wide range of temperature (294-423 K) and NaCl concentration (1-6 M) was performed through MD simulation, and a thorough data set was provided. In addition, the role of pressure variation and composition of salts on the $CO_2$ diffusion coefficient changes were examined. Finally, the direct numerical simulation was employed to highlight differences in the $CO_2$ dissolution process between assuming the $CO_2$ diffusion coefficient in brine rather than pure water. The highlight important points of this research are as follows:

1- The $CO_2$ diffusion coefficient decreases monotonically with increasing NaCl salinity concentration at a constant temperature, while the diffusion coefficients go through the trend of rising as the temperature increases for each concentration. The formation of $Na^+$ cation hydration shells, which act as an obstacle to $CO_2$ movement, is the primary reason for the diffusion coefficient reduction. Since fewer water molecules are present in the hydration shell of cations at higher temperatures, a higher diffusion coefficient also results. Also, for predicting the $CO_2$ diffusion coefficient, a new correlation was proposed in terms of temperature and NaCl salinity concentration with the relative accuracy of $R^2 = 0.988$.

2- Pressure variation has a negligible influence on the $CO_2$ diffusion coefficient in water/saline brine under different temperature and salinity concentrations. While salinity compositions change the $CO_2$ diffusion coefficient, in which $CaCl_2$ and KCl have the highest and lowest impact on the $CO_2$ diffusion coefficient reduction, respectively. It is worth noting that there exists not an accurate relationship between the type of salt and the $CO_2$ diffusion coefficient decrease.

3- Salts can lessen the value of the $CO_2$ diffusion coefficient following three possible mechanisms, including restriction of the $CO_2$ movement via direct interplays of $CO_2$ molecules and cations, interrupting the $CO_2$ freedom through the formation of cations' hydration, and reducing the available space and creation of $CO_2$ movement barrier by the appearance of ionic aggregations.

4- According to direct numerical simulation results, the lack of data for the $CO_2$ diffusion coefficient in brine results in false predictions and substantial errors in calculating the $CO_2$ dissolution behavior. This includes overestimating or underestimating dissolved $CO_2$ dissolution flux and the onset of different $CO_2$ dissolution regimes.

Since there is lacking data for the $CO_2$ diffusion coefficient in water/saline brine under a real geological condition, we hope that our research paves the way for future research pertaining to $CO_2$ sequestration in saline aquifers so as to have long-term and safe underground $CO_2$ storage.

# Supporting Information "A Molecular Dynamics Study on $CO_2$ Diffusion Coefficient in Saline Water in a Wide Range of Temperatures, Pressures, and Salinity Concentrations: Implications to $CO_2$ Geological Storage"

Sina Omrani[a,‡] , Mehdi Ghasemi[b,‡], Saeed Mahmoodpour[c, *], Ali Shafiei[a,*] , Behzad Rostami[b]

[a] *Institute of Petroleum Engineering, College of Engineering, University of Tehran, Tehran, Iran.*

[b] *Petroleum Engineering Program, School of Mining & Geosciences, Nazarbayev University, Nur-Sultan, Astana 010000, Kazakhstan*

[c] *Institute of Applied Geosciences, Geothermal Science and Technology, Technische Universität Darmstadt, Darmstadt, Germany*

*‡ S. Omrani and M.Ghasemi contributed equally to this work.*

** Corresponding author*

*E-mail address: ali.shafiei@nu.kz.edu (A. Shafiei), saeed.mahmoodpour@tu-darmstadt.de (S. Mahmoodpour)*

# 1. Literature Review

In this section, we present a thorough review of previous studies on the $CO_2$ diffusion coefficient in water/brine. As time passed by, the advancement of technologies helped researchers to widen the range of their measurements. Earlier works were limited to a narrow range of temperature and pressure. Furthermore, most of them were performed in pure water conditions. In this respect, various methods were applied, such as the Pressure Decay method [1-3], Taylor Dispersion method [4], Raman Spectroscopy [5, 6], Microfluidic method [7], Nuclear Magnetic Resonance (NMR) [8], Molecular dynamics (MD) simulation [9, 10], etc. **Table S4** lists these studies. We pointed out some of them with more novelty in the original manuscript.

**Table S4-** Previous research outcomes on $CO_2$ diffusion coefficient measurement in water/brine. Note that T and P are temperature and pressure, respectively.

| Year | Researcher | Solution (g/L) | T (K) | P (MPa) | Diffusion Coefficient ($\times 10^{-9} \frac{m^2}{s}$) |
|---|---|---|---|---|---|
| 1956 | Scriven et al. [11] | pure water | 298 | - | 1.87 |
| 1961 | Woods et al. [12] | pure water | 298 | - | 1.95 |
| 1964 | Tang and Himmelblau [13] | pure water | 298 | 0.1 | 1.92 |
| 1964 | Vivian and King [14] | pure water | 298 | - | 2 |
| 1964 | Unver and Himmelblau [15] | pure water | 298 | 0.1 | 1.85 |
| 1965 | Thomas and Adams [16] | pure water | 298 | 0.1 | 1.95 |
| 1967 | Ferrell and Himmelblau [17] | pure water | 298 | - | 1.92 |
| 1988 | Renner [3] | brine | 311 | 1.544-5.832 | 1.92 |
| 1994 | Tamimi et al. [18] | pure water | 293-368 | 1 | 1.76-8.20 |
| 1996 | Frank et al. [19] | pure water | 298-328 | 1 | 1.97-3.67 |
| 1996 | Wang et al. [20] | brine | 311 | 1.524-5.178 | 2.925-4.827 |
| 1997 | Hirai et al. [21] | pure water | 286 | 29.4, 39.2 | 1.35, 1.45 |
| 2005 | Tewes and Boury [22] | pure water | 313 | 3.0-9.0 | 0.47-1.8 |
| 2006 | Yang et al. [23] | brine | 300, 330 | 1.786-5.647 | 170.7-269.8 |
| 2009 | Farajzadeh et al. [24] | brine | 303 | 1.0-5.0 | 2-280 |
| 2012 | Nazari Moghaddam et al. [25] | pure water | 298 | 2.185-5.861 | 9.07-9.86 |
| 2012 | Sell et al. [7] | pure water, brine | 299 | 0.5-5.0 | 1.86, 0.55-1.4 |
| 2012 | Garcia-Rates et al. [9] | brine | 333-453 | 5.0-50.0 | 2.4-12.55 |
| 2013 | Azin et al. [2] | brine | 305-325 | 5.9-6.9 | 3.52-6.16 |
| 2013 | Lu et al. [5] | pure water | 268-473 | 10.0-45.0 | 0.76-16.1 |
| 2013 | Wang et al. [26] | brine | 423 | 3.43-8.02 | 238.35-251.34 |
| 2014 | Cadogan et al. [4] | pure water | 298-423 | 14.2-49.3 | 2.218-12.33 |
| 2014 | Moultos et al. [10] | pure water | 298-623 | 0.1-100.0 | 2-55 |
| 2015 | Zhang et al. [27] | brine | 298-343 | 0.5-1.17 | 1.2-1.91 |
| 2015 | Cadogan et al. [8] | pure water, brine | 298 | 0.1 | 1.25-2.13 |
| 2015 | Belgodere et al. [6] | brine | 294 | 4.0 | 0.92-1.71 |
| 2015 | Raad et al. [2] | pure water, brine | 303, 313 | 5.880-6.265 | 0.678-23.3 |
| 2017 | Zarghami et al. [28] | brine | 323-348 | 1.745 | 3.6-8.2 |
| 2018 | Shi et al. [29] | brine | 323 | 6.0 | 1.25-82 |
| 2021 | Li et al. [30] | brine | 313-373 | 8.28-30.94 | 0.0166-0.0961 |

## 2. Accuracy Evaluation of Combined Force Fields

At the first stage, due to the existence of various models for $CO_2$ and NaCl, MD simulations have been carried out to determine the most accurate combined model. After some preliminary assessment, TraPPE [31], TraPPE (Flex) [32], EPM2 [33] models were considered for $CO_2$, while models developed by Smith et al. [34] and Joung and Cheatham [35] were regarded for NaCl. Moreover, according to the literature, the SPC/E model was assumed for water molecules [36]. In summary, 6 different combined models, namely TraPPE-SPC/E-Smith, EPM2-SPC/E-Smith, TraPPE-SPC/E-JC, EPM2-SPC/E-JC, TraPPE (Flex)-SPC/E-Smith, and TraPPE (Flex)-SPC/E-JC, were selected to compare the result of the self-diffusion coefficient obtained by MD simulations with available experimental data measured by Belgodere et al. [6] in distinct NaCl molarities of 1, 2, and 3. To evaluate and compare the accuracy of combined models with the measured values in a laboratory, various statistical analyses, including average percent relative error (APRE), average percent absolute relative error (AAPRE), standard deviation (SD) of error, and root mean square error (RSME) and coefficient of determination ($R^2$) [37]. The average value of the self-diffusion coefficient (3 times repetition) for different molarities is shown in **Table S5**. It should be noted that the employed simulation procedure was explained in the original manuscript. Also, the results of statistical analyses are revealed in

**Table S6**. As is clear, the most accurate model is TraPPE (Flex)-SPC/E-JC, which was adopted for the rest of the simulations.

**Table S5**- The average value of the self-diffusion coefficient for 6 different combined models. Note that (A) TraPPE-SPC/E-Smith, (B) EPM2-SPC/E-Smith, (C) TraPPE-SPC/E-JC, (D) EPM2-SPC/E-JC, (E)TraPPE (Flex)-SPC/E-Smith, and (F) TraPPE (Flex)-SPC/E-JC

| | **1 M** | **2 M** | **3 M** |
|---|---|---|---|
| **Ref.** [6] | 1.5500 | 1.3800 | 1.2900 |
| **A** | 1.8440 | 1.6326 | 1.2940 |
| **B** | 1.6966 | 1.6033 | 1.3757 |
| **C** | 1.7415 | 1.6080 | 1.2080 |
| **D** | 1.7720 | 1.5553 | 1.1500 |
| **E** | 1.7122 | 1.5812 | 1.2780 |
| **F** | 1.6831 | 1.5164 | 1.3210 |

**Table S6**- Statistical error analysis of 6 different combined models. Note that (A) TraPPE-SPC/E-Smith, (B) EPM2-SPC/E-Smith, (C) TraPPE-SPC/E-JC, (D) EPM2-SPC/E-JC, (E) TraPPE (Flex)-SPC/E-Smith, and (F) TraPPE (Flex)-SPC/E-JC.

| **Analyses** | **Molarity** | **A** | **B** | **C** | **D** | **E** | **F** |
|---|---|---|---|---|---|---|---|
| APRE | 1 | −18.9677 | −9.4623 | −12.3548 | −14.3225 | −9.6312 | −9.3322 |
| | 2 | −18.3091 | −9.4623 | −16.5217 | −13.7053 | −14.5246 | −9.2864 |
| | 3 | −0.3100 | −6.6149 | 6.3565 | 7.8527 | −0.8123 | −1.0542 |
| | T | −12.5289 | −8.5132 | −7.5066 | −6.7250 | −8.3227 | −6.5576 |
| AAPRE | 1 | 0.2940 | 0.1466 | 0.1915 | 0.2220 | 0.1748 | 0.1546 |
| | 2 | 0.2526 | 0.2233 | 0.2280 | 0.1753 | 0.2230 | 0.1842 |
| | 3 | 0.0580 | 0.0853 | 0.1080 | 0.1400 | 0.0430 | 0.0640 |
| | T | 0.2015 | 0.1517 | 0.1758 | 0.7911 | 0.1470 | 0.1343 |
| RMSE | 1 | 0.3170 | 0.1760 | 0.2002 | 0.2421 | 0.1866 | 0.1925 |
| | 2 | 0.2665 | 0.2350 | 0.2400 | 0.2102 | 0.2125 | 0.2232 |
| | 3 | 0.0687 | 0.0901 | 0.1356 | 0.1568 | 0.0801 | 0.0542 |
| | T | 0.2174 | 0.1670 | 0.1919 | 0.2030 | 0.1597 | 0.1598 |
| SD | 1 | 0.2505 | 0.1391 | 0.1826 | 0.1913 | 0.1745 | 0.2151 |
| | 2 | 0.2365 | 0.2485 | 0.2459 | 0.1865 | 0.1945 | 0.2066 |
| | 3 | 0.0652 | 0.0855 | 0.1486 | 0.1489 | 0.0756 | 0.0342 |
| | T | 0.1841 | 0.1577 | 0.1924 | 0.1756 | 0.1482 | 0.1520 |
| $R^2$ | 1 | 0.9135 | 0.9784 | 0.9633 | 0.9457 | 0.9735 | 0.9749 |
| | 2 | 0.9361 | 0.9501 | 0.9480 | 0.9893 | 0.9794 | 0.9814 |
| | 3 | 0.9999 | 0.9927 | 0.9932 | 0.9867 | 0.9897 | 0.9981 |
| | T | 0.9499 | 0.9737 | 0.9682 | 0.9739 | 0.9809 | 0.9848 |

# 3. Assessment of the Effect of Box Size on the Diffusion Coefficient

Various researchers have investigated the finite-size effect of the diffusion coefficient computed by molecular dynamics simulations [38, 39]. A common method of checking the effect of box size is to run simulations in boxes of different lengths and measure the difference between them [10]. The objective is to measure the diffusion coefficient at the thermodynamic limit, or, in other words, to compute diffusion coefficients in an infinite box ($1/L \rightarrow 0$). Three simulation box sizes were considered: 3, 4.5, and 7 nm. **Figure S14** shows the results. As it can be seen, compared to each other, there is not much of a difference.

This fact is confirmed by calculating the Yeh–Hummer correction term [40]. It also shows that the required correction is minor (almost two orders of magnitude lower than the diffusion coefficient).

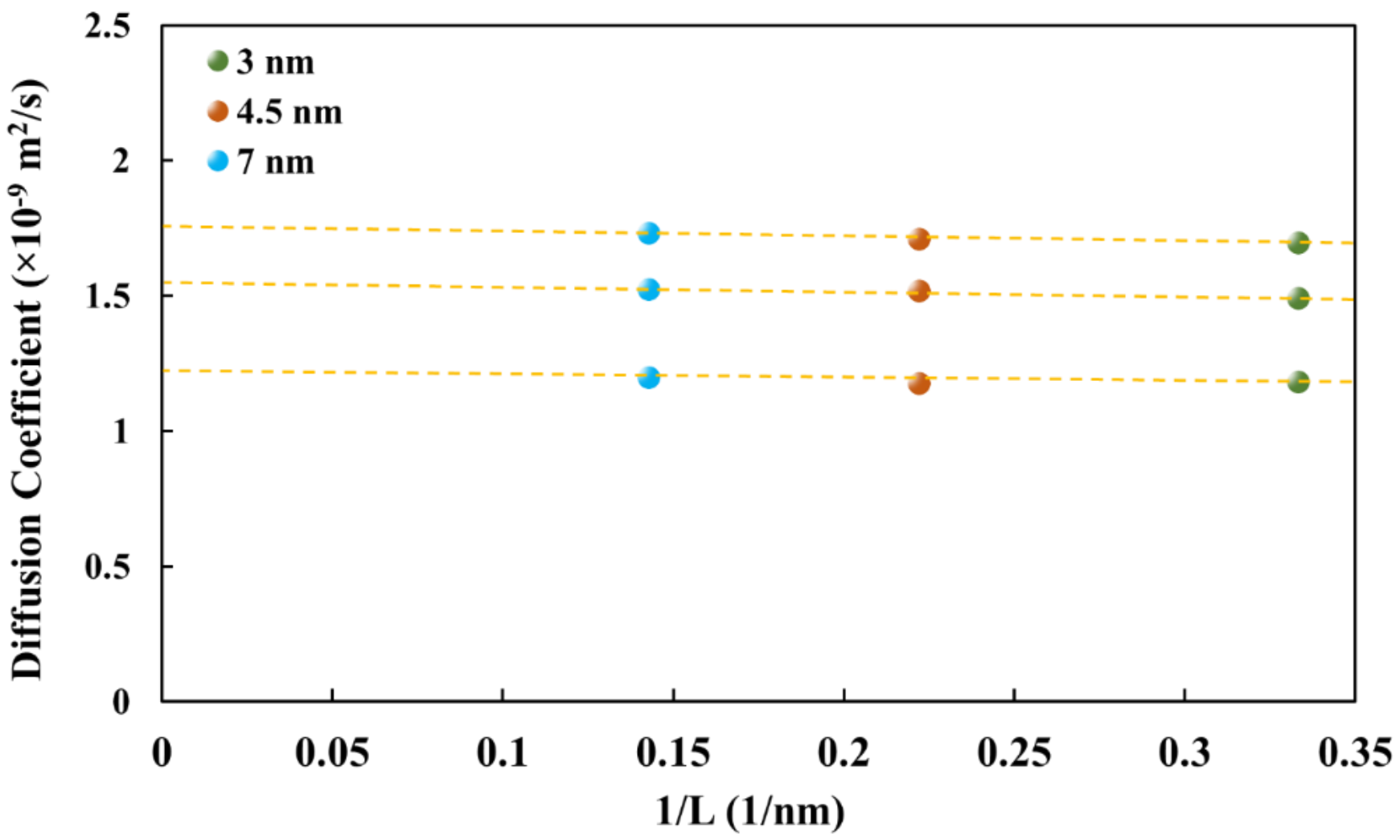

**Figure S14.** The $CO_2$ diffusion coefficient at 294 K and 10 MPa in the box with the length of 3 (green), 4.5 (Orange), and 7 (blue) nm.

# 4. Results of RDFs

The RDFs of all investigated pairs are shown hereunder. However, only two cases of 294 K and 423 K are shown in the original manuscript.

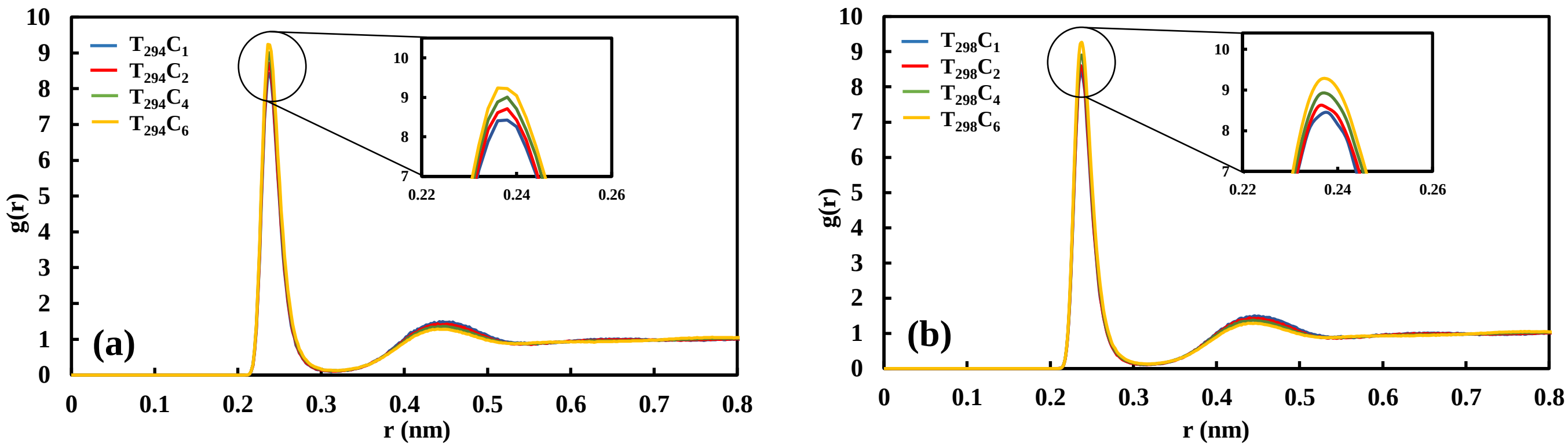

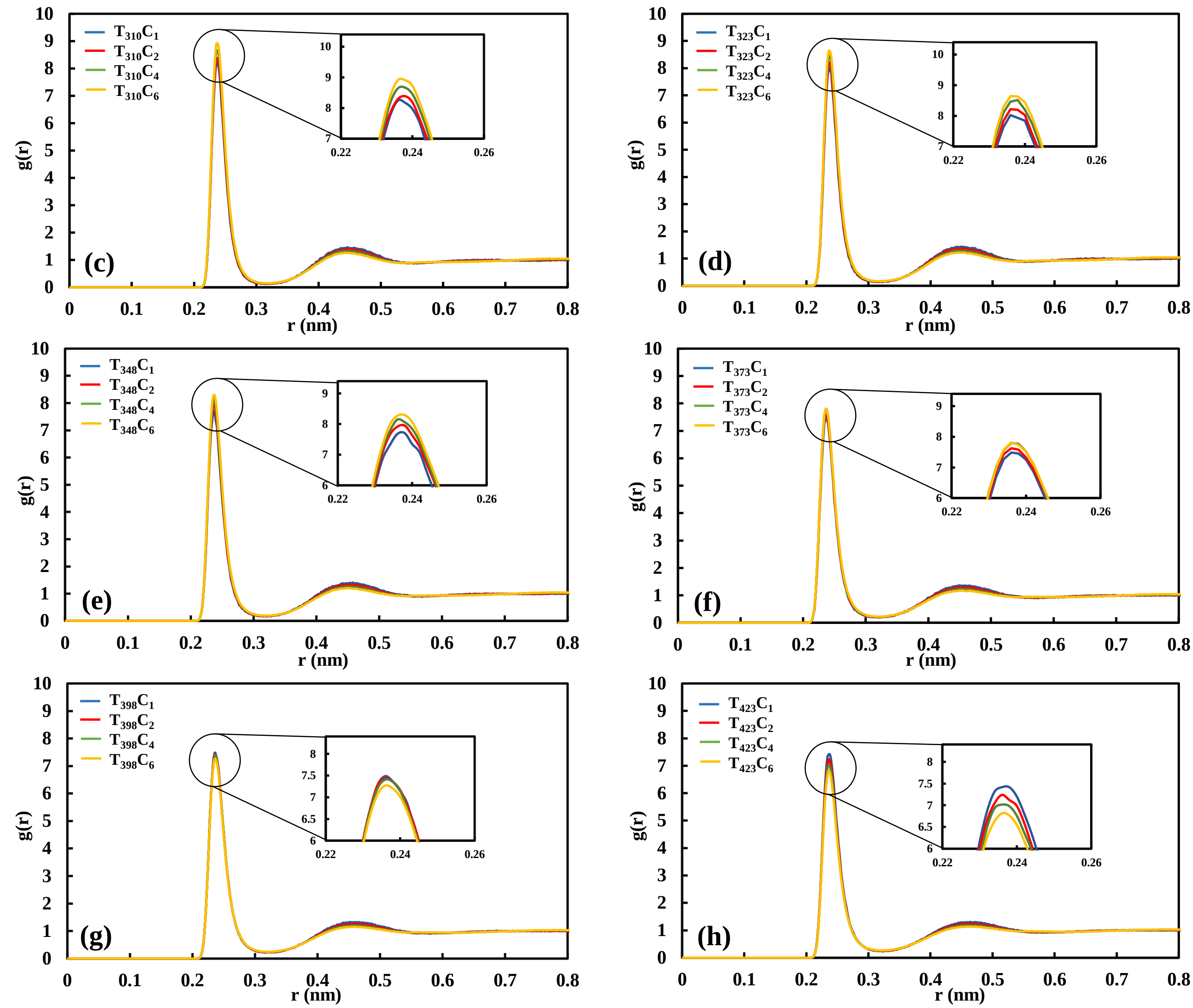

**Figure S15-** RDF curves of $Na\text{-}O_{H_2O}$ in various NaCl solutions at the temperatures of (a) 294 K, (b) 298 K, (c) 310 K, (d) 323 K, (e) 348 K, (f) 373, (g) 398 K, and (h) 423 K.

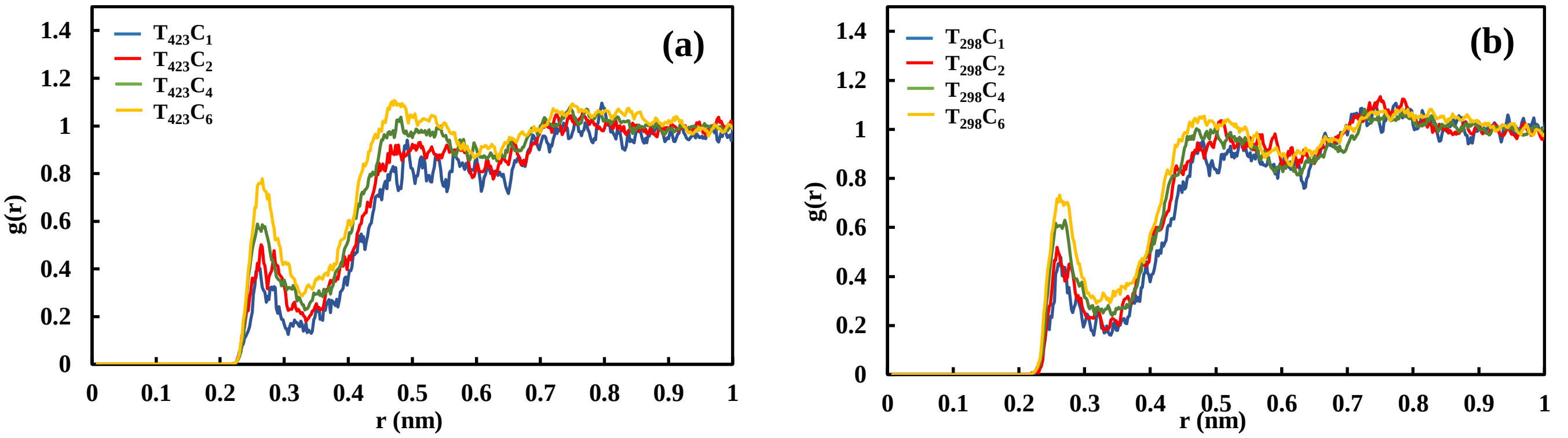

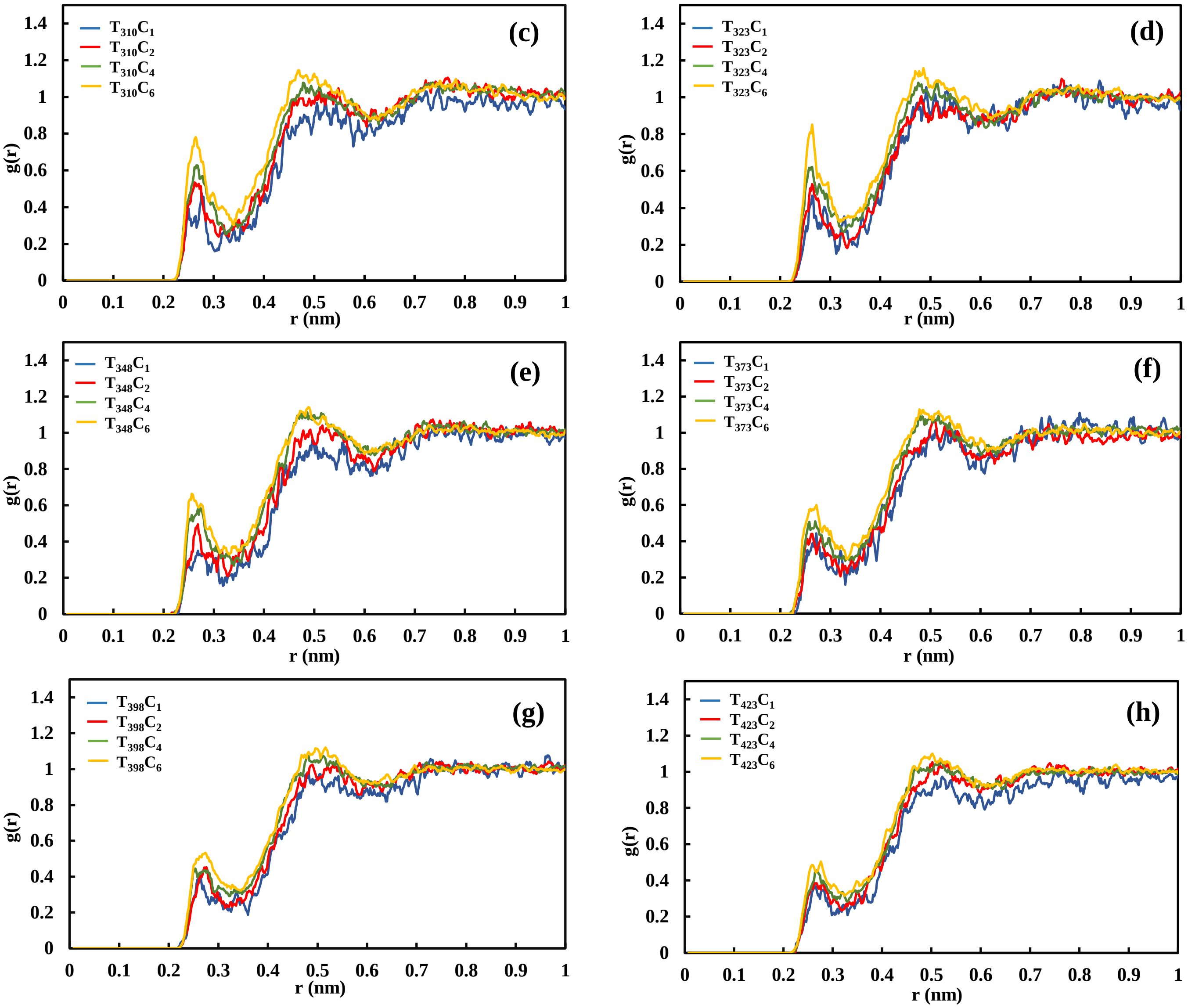

**Figure S16**- RDF curves of $Na\text{-}O_{CO_2}$ in various NaCl solutions at the temperatures of (a) 294 K, (b) 298 K, (c) 310 K, (d) 323 K, (e) 348 K,( f) 373 K, (g) 398 K, and (h) 423 K.

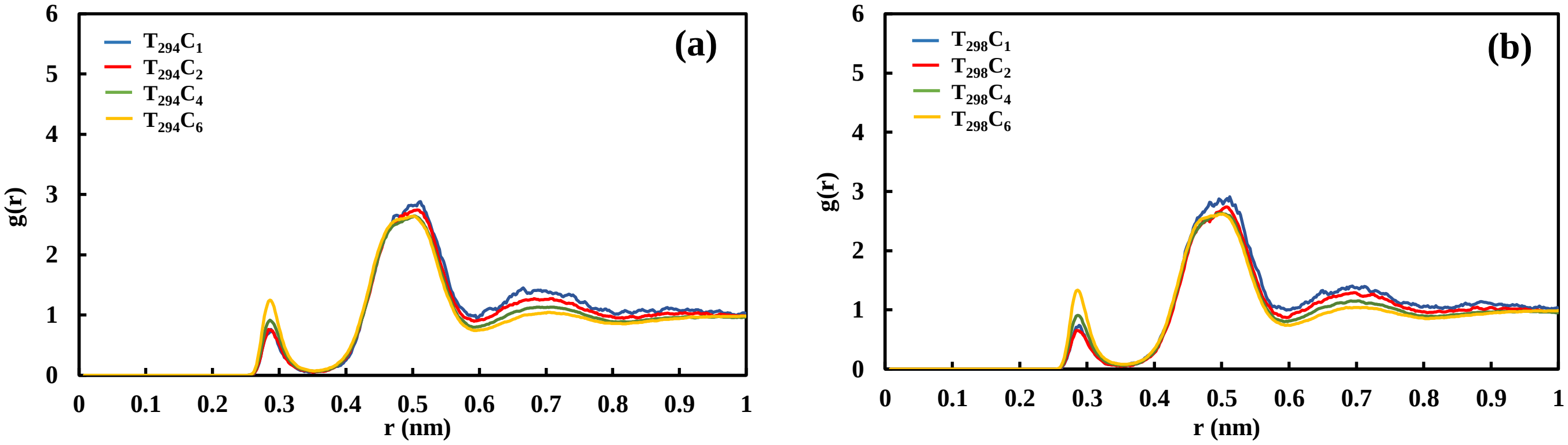

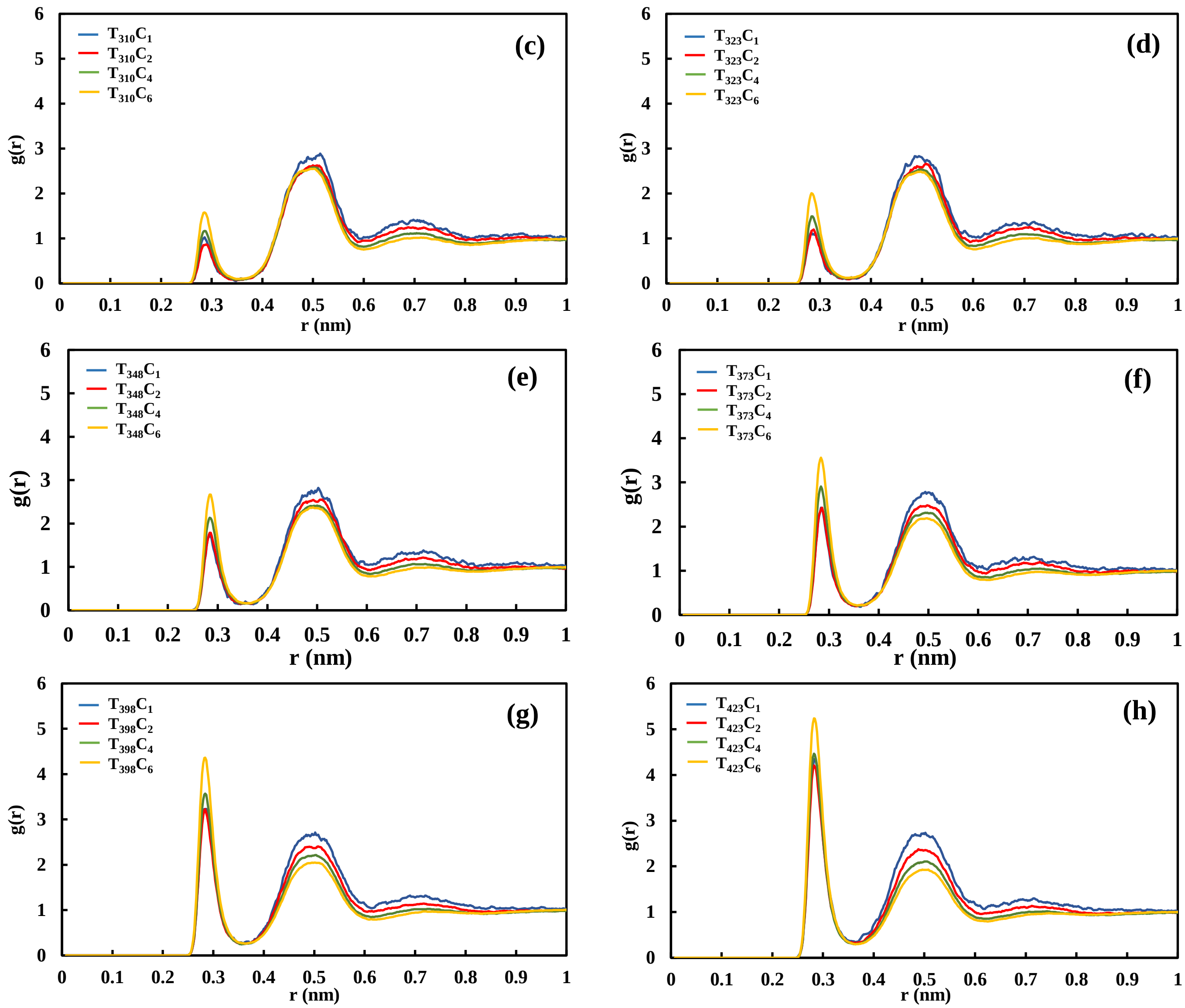

**Figure S17**- RDF curves of $Na\text{-}Cl$ in various NaCl solutions at the temperatures of (a) 294 K, (b) 298 K, (c) 310 K, (d) 323 K, (e) 348 K, (f) 373 K, (g) 398 K, and (h) 423 K.

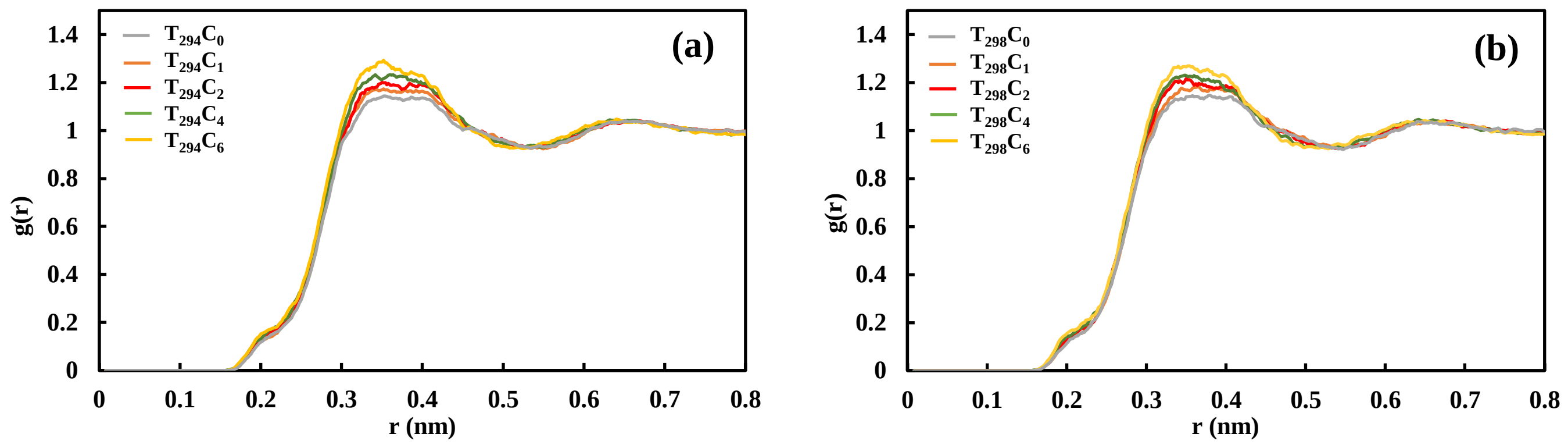

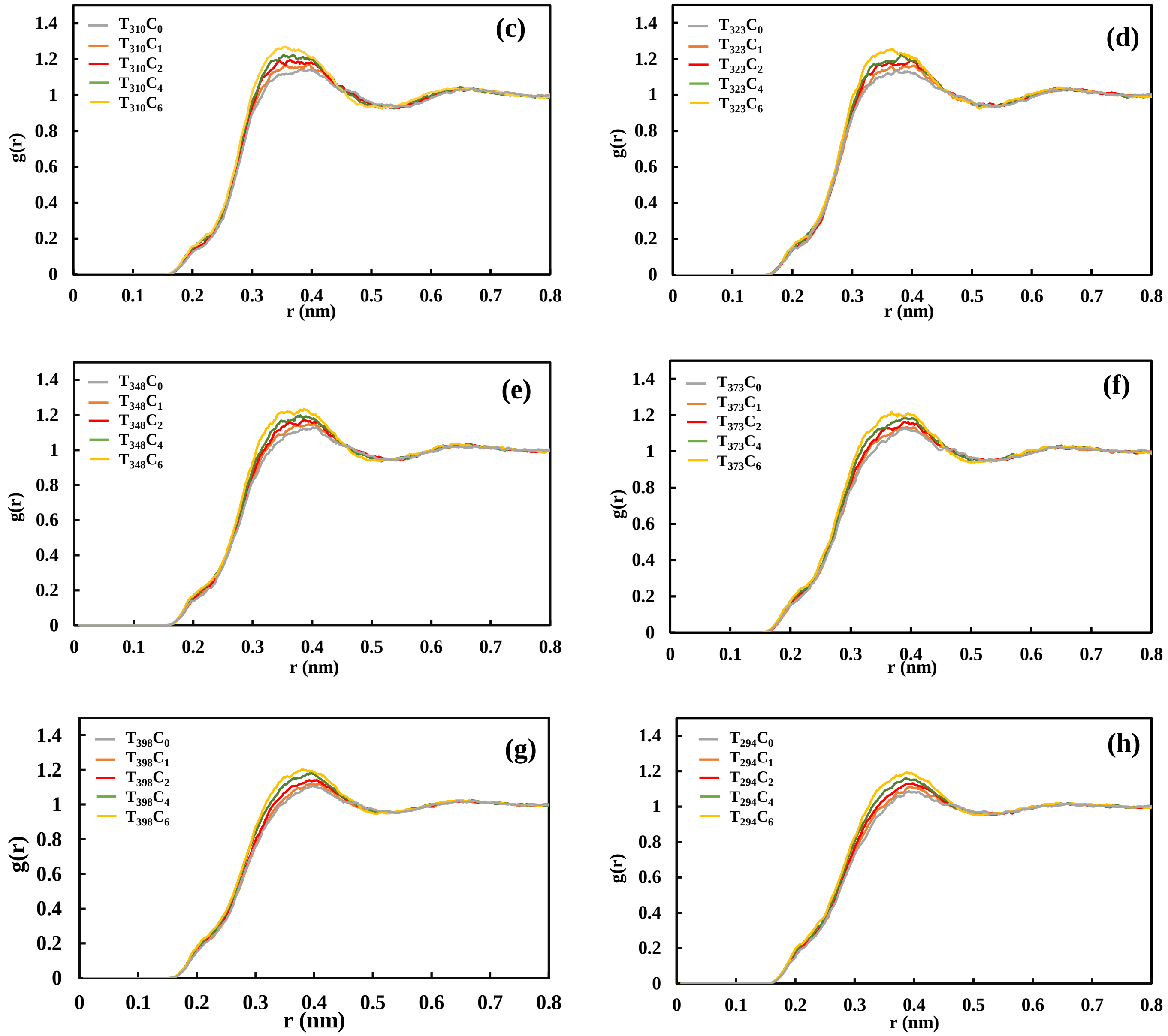

**Figure S18**-RDF curves of $O_{CO_2}$-$H_{H_2O}$ in various aqueous solutions at the temperatures of (a) 294 K, (b) 298 K, (c) 310 K, (d) 323 K, (e) 348 K, (f) 373 K, (g) 398 K, and (h) 423 K.

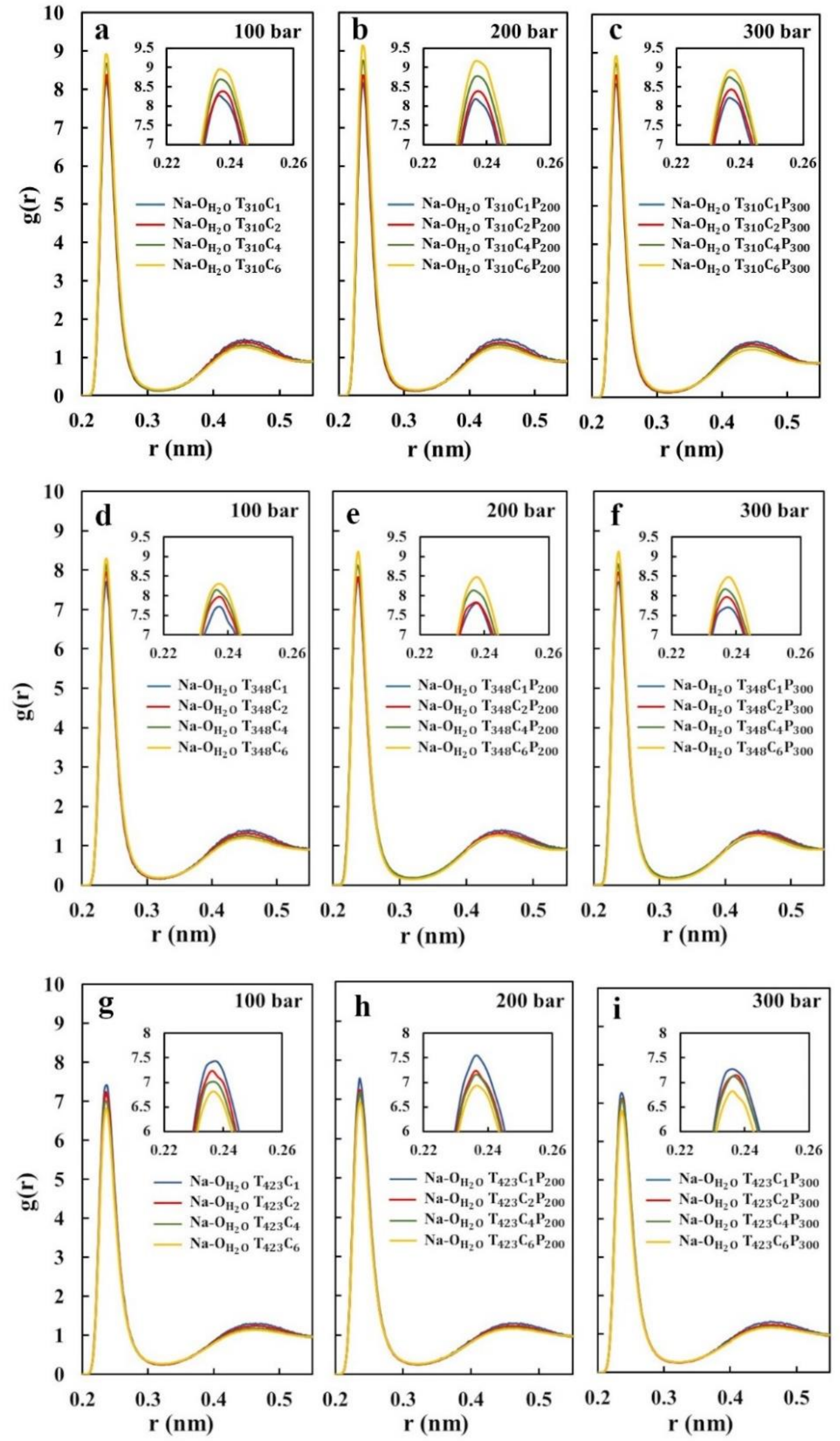

**Figure S19-** RDF values between Na-$O_{H_2O}$ at 310 K and pressures of a) 100 bar b) 200 bar c) 300 bar, at 348 K and pressures of d) 100 bar, e) 200 bar, and f) 300 bar, and at 423 K and pressures of g) 100 bar, h) 200 bar, and i) 300 bar.

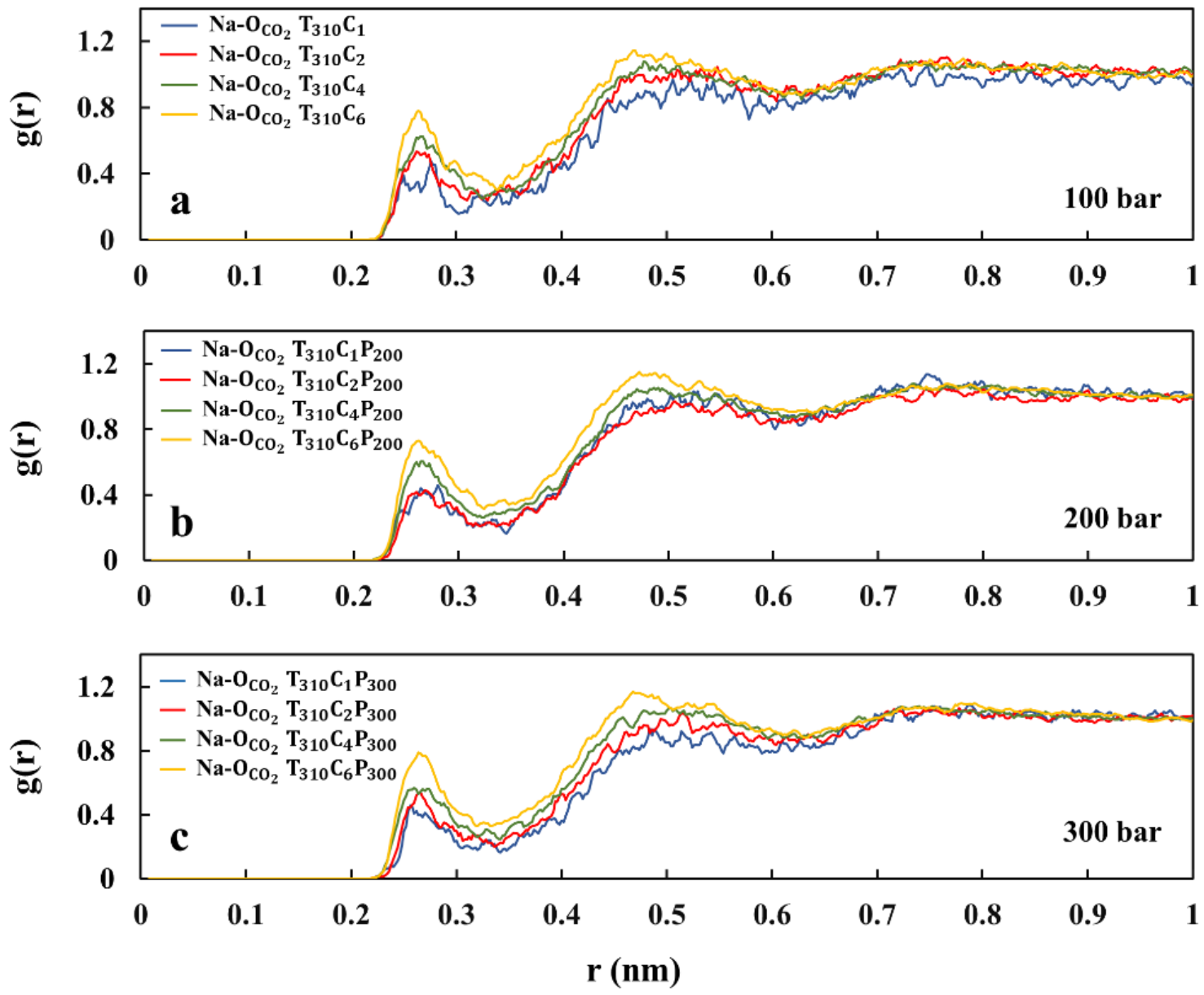

**Figure S20**- RDF values between Na-$O_{CO_2}$ at 310 K and pressures of a) 100 bar b) 200 bar c) 300 bar.

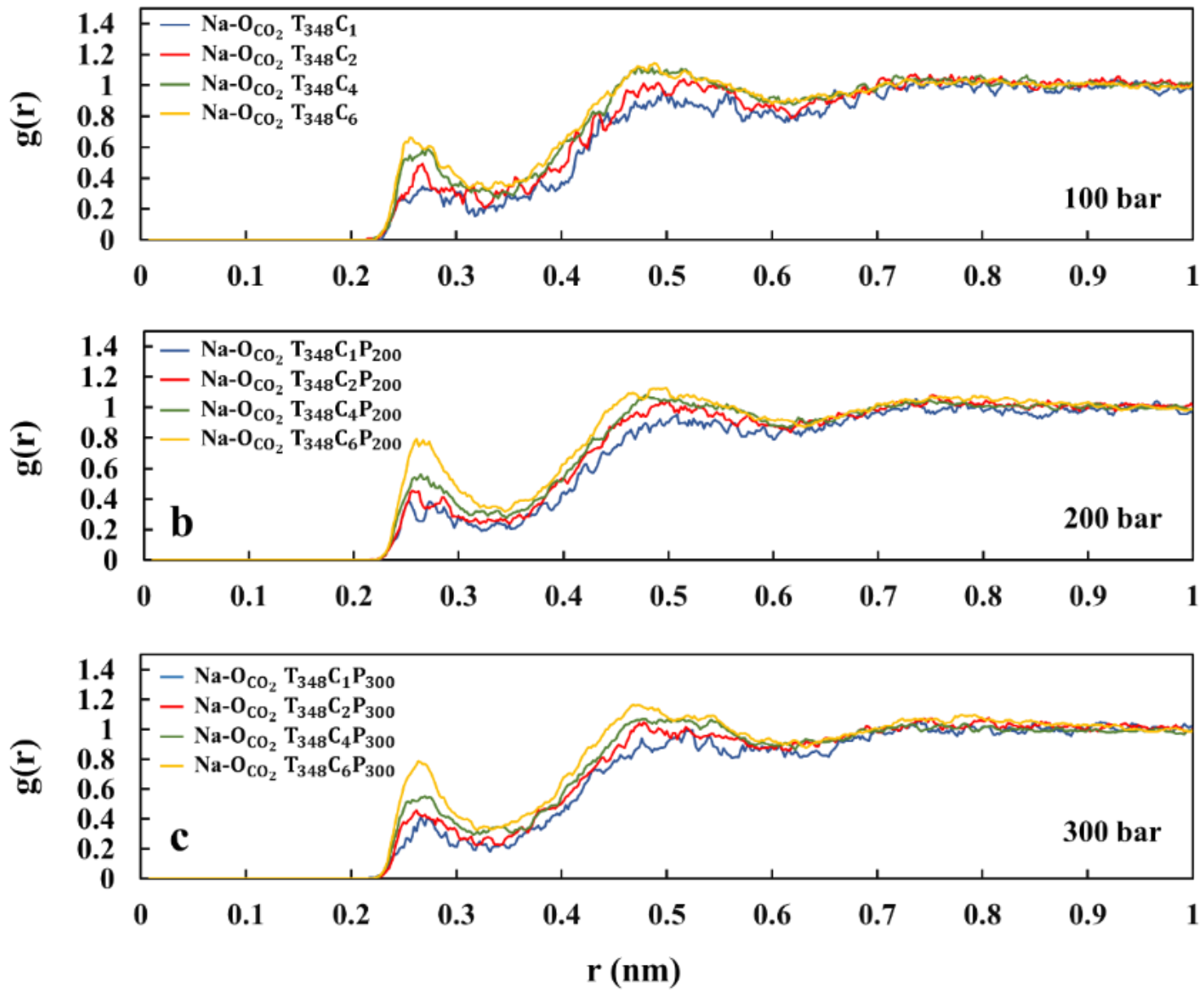

**Figure S21**- RDF values between Na-$O_{CO_2}$ at 348 K and pressures of a) 100 bar, b) 200 bar, and c) 300 bar.

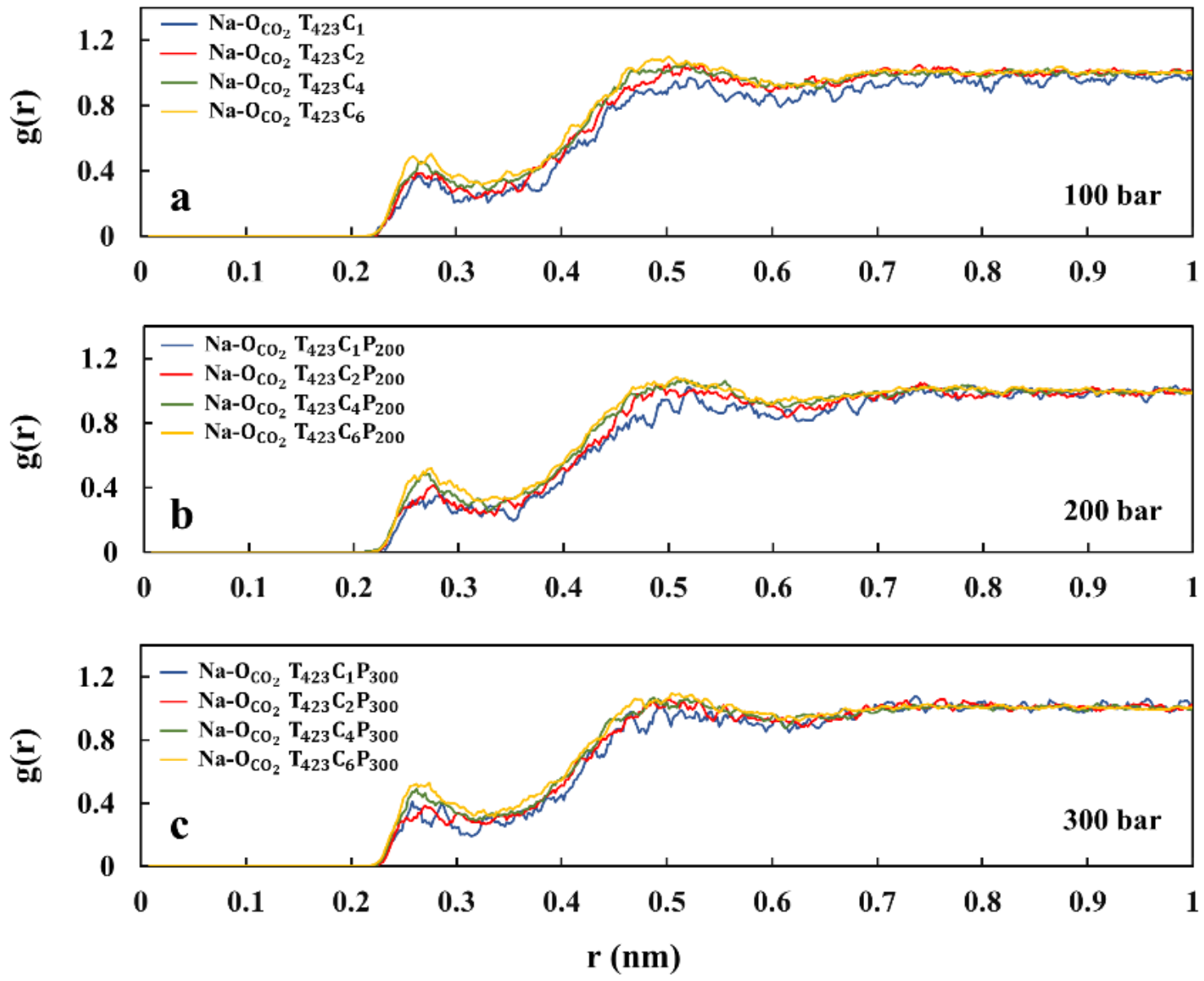

**Figure S22**- RDF values between Na-$O_{CO_2}$ at 423 K and pressure of a) 100 bar, b) 200 bar, and c) 300 bar.

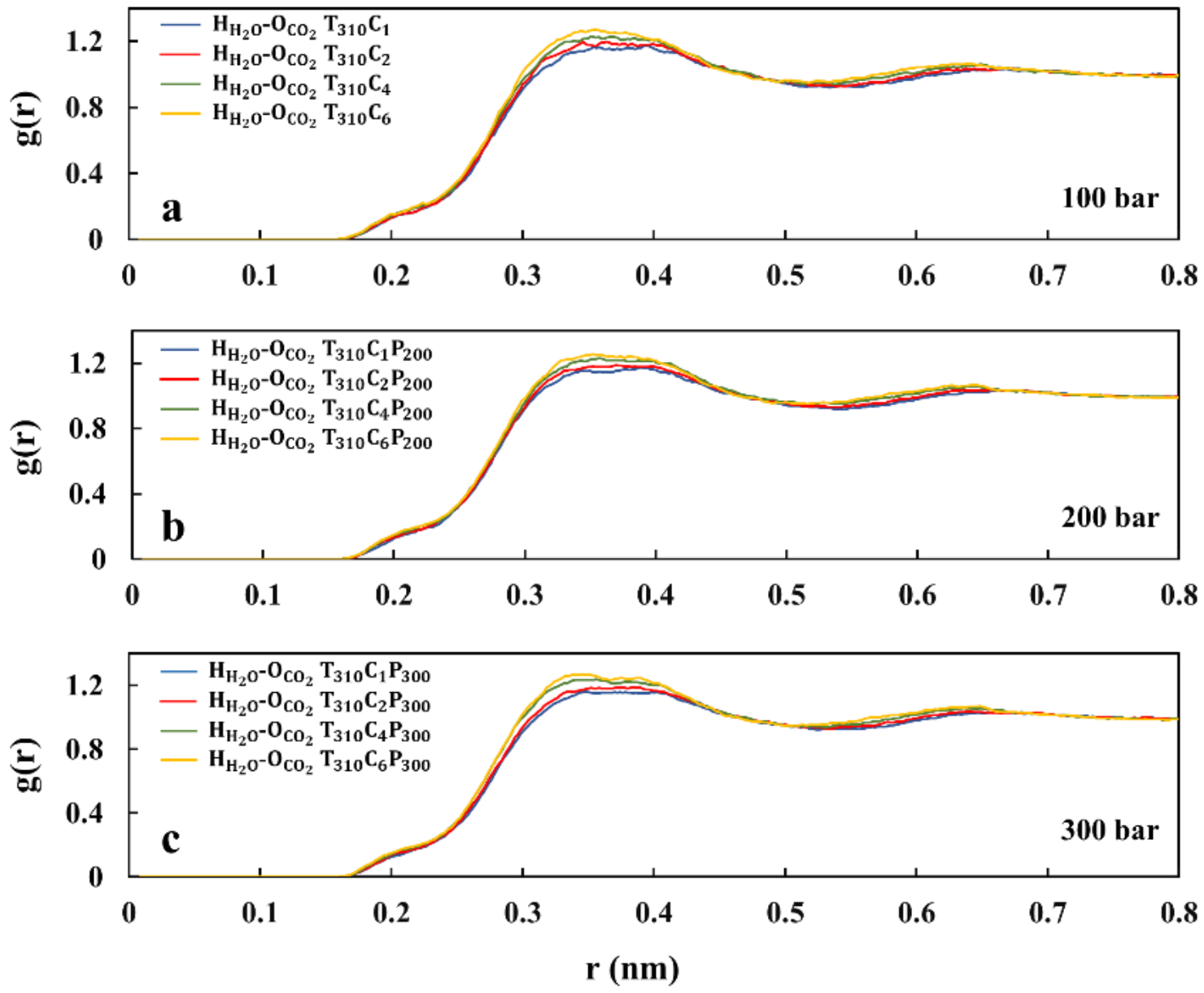

**Figure S23**- RDF values between $H_{H_2O}$-$O_{CO_2}$ at 310 K and pressure of a) 100 bar, b) 200 bar, and c) 300 bar

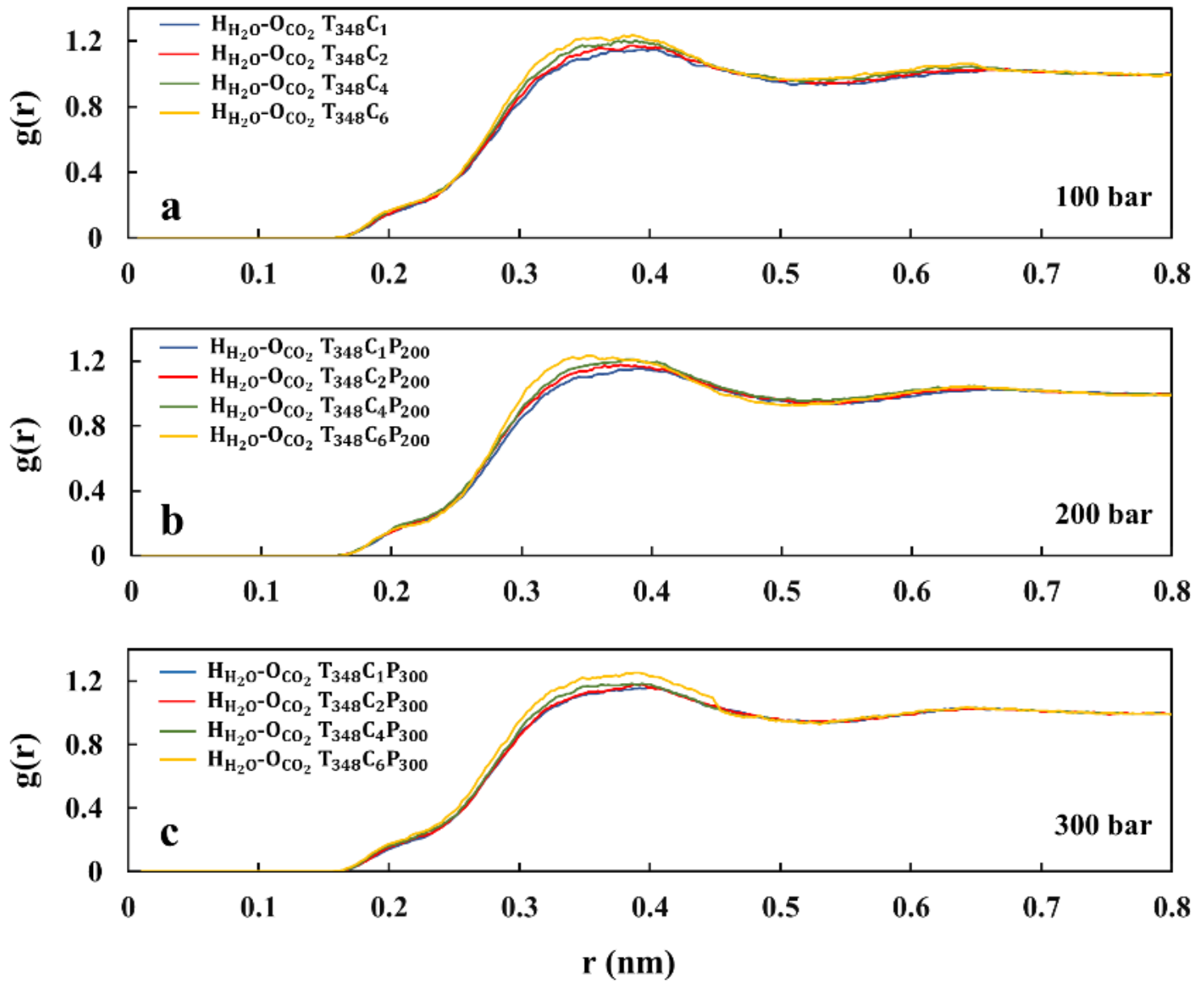

**Figure S24**- RDF values between $H_{H_2O}$-$O_{CO_2}$ at 348 K and pressure of a) 100 bar, b) 200 bar, and c) 300 bar

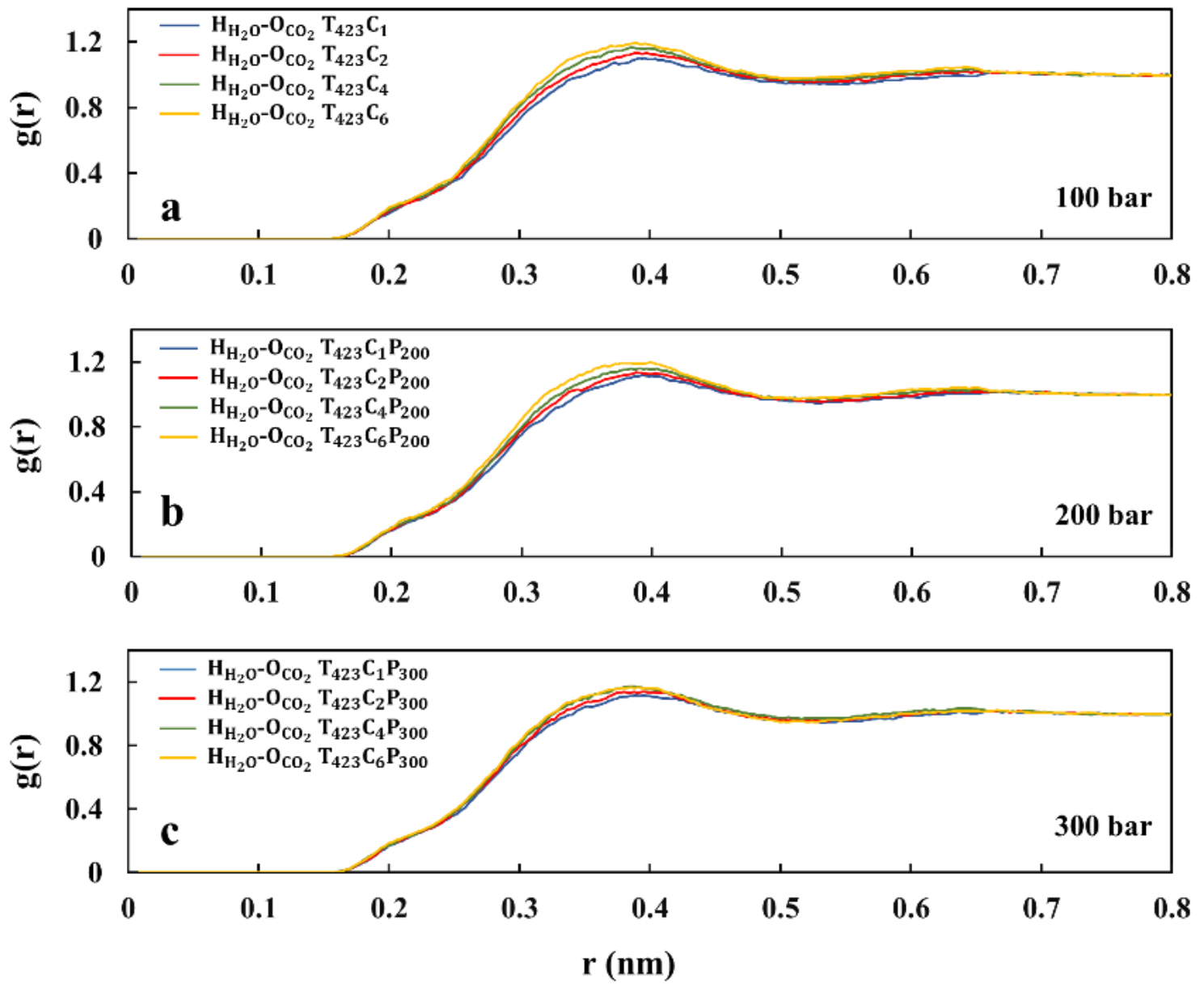

**Figure 25**- RDF values between $H_{H_2O}$-$O_{CO_2}$at 423 K and pressure of a) 100 bar, b) 200 bar, and c) 300 bar

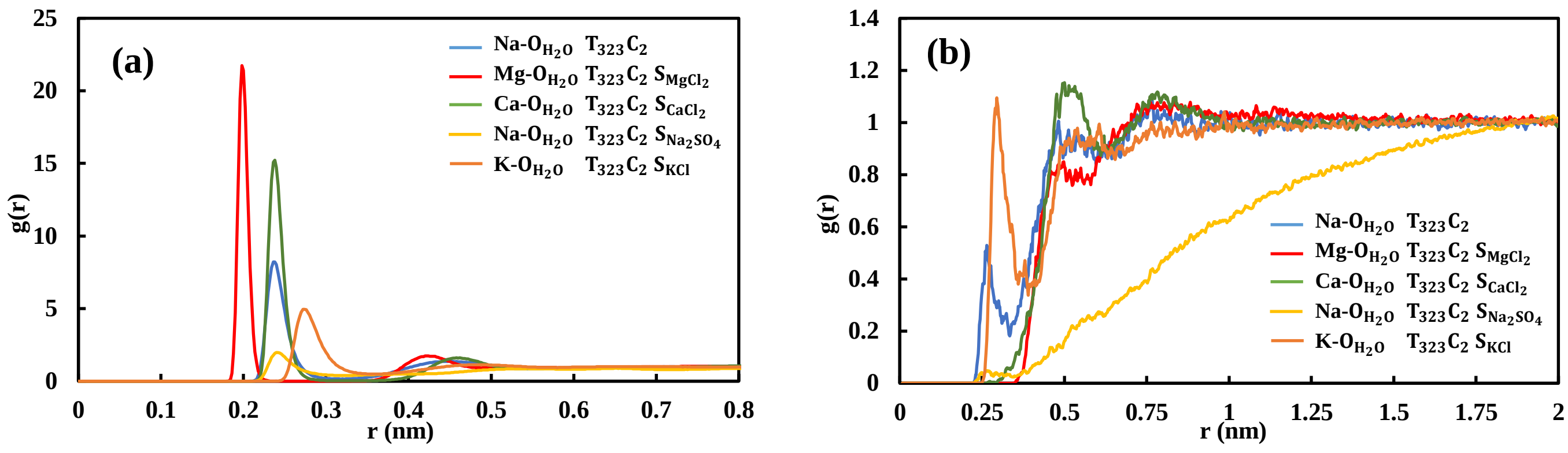

**Figure S26**- The RDF values between cations and (a) $O_{H_2O}$ and (b) $O_{CO_2}$ in various saline solutions at the temperatures 323 K and 2 M concentration.

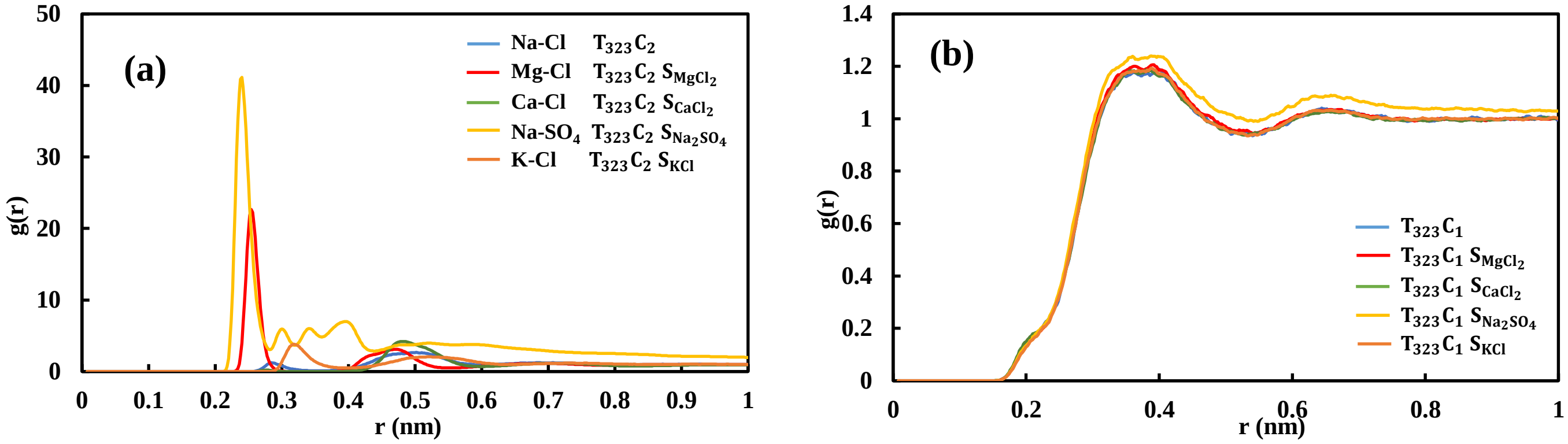

**Figure S27**- The RDF values between (a) cations-anions and (b) $H_{H_2O}$-$O_{H_2O}$ in various saline solutions at the temperatures 323 K and 2 M concentration.